\documentclass[a4paper,fleqn]{cas-sc}

\pdfoutput=1

\usepackage{multirow}
\usepackage[authoryear,longnamesfirst]{natbib}
\usepackage{CJKutf8}
\usepackage{tabularray}
\usepackage{tabularx}
\usepackage{booktabs}
\usepackage[normalem]{ulem}
\usepackage{float} 
\usepackage{caption} 
\usepackage{subcaption} 
\usepackage{tikz}
\usepackage[super]{nth}
\usepackage{varwidth}  
\usepackage{mdframed} 
\usepackage{soul}
\usepackage[detect-weight=true,group-separator={,},
            output-decimal-marker={.}]{siunitx}
\newcommand{\noop}[1]{}
\begin{document}
\let\WriteBookmarks\relax
\def\floatpagepagefraction{1}
\def\textpagefraction{.001}
\def\code#1{\texttt{#1}}

\shorttitle{Classification of Quality Characteristics in Online User Feedback}

\shortauthors{EC Groen et~al.}

\title [mode = title]{Classification of Quality Characteristics in Online User Feedback using Linguistic Analysis, Crowdsourcing and LLMs} 

\author
[1,2]
{Eduard C. Groen}
[type=,
auid=,
bioid=1,
prefix=,
role=,
orcid=0000-0002-5551-8152]
\cormark[1] 
\ead{eduard.groen@iese.fraunhofer.de} 
%
\credit{Conceptualization, methodology, validation, formal analysis, investigation, data curation, writing -- original draft, visualization}

\affiliation[1]{organization={Fraunhofer IESE},
    addressline={Fraunhofer-Platz 1}, 
    postcode={67663}, 
    postcodesep={}, 
    city={Kaiserslautern},
    country={Germany}}

\affiliation[2]{organization={Utrecht University, Department of Information and Computing Sciences},
    addressline={Princetonplein 5}, 
    postcode={3584 CC}, 
    postcodesep={}, 
    city={Utrecht}, 
    country={The Netherlands}}

\author
[2]
{Fabiano Dalpiaz}
[type=,
auid=,
bioid=2,
prefix=,
role=,
orcid=0000-0003-4480-3887]
\cormark[1] 
\ead{f.dalpiaz@uu.nl} 
%
\credit{Conceptualization, methodology, validation, investigation, writing -- review \& editing, supervision}

\author
[2]
{Martijn {van Vliet}}
[type=,
auid=,
bioid=3,
prefix=,
role=,
orcid=]
\ead{m.vanvliet@uu.nl} 
%
\credit{Methodology for RQ2, validation for RQ2, investigation for RQ2, writing -- original draft for RQ2}

\author
[2]
{Boris Winter}
[type=,
auid=,
bioid=4,
prefix=,
role=,
orcid=]
\ead{boris@prostrive.io} 
%
\credit{Validation for RQ2, investigation for RQ2, writing -- original draft for RQ2}

\author
[1,3]
{Joerg Doerr}
[type=,
auid=,
bioid=5,
prefix=,
role=,
orcid=0000-0003-0048-399X]
\ead{joerg.doerr@iese.fraunhofer.de} 
%
\credit{Conceptualization, methodology for RQ1, writing -- original draft for RQ1, writing -- review \& editing}

\affiliation[3]{organization={University Kaiserslautern--Landau},
    addressline={Gottlieb-Daimler-Stra{\ss}e}, 
    postcode={67663}, 
    postcodesep={}, 
    city={Kaiserslautern},
    country={Germany}}

\author
[2]
{Sjaak Brinkkemper}
[type=,
auid=,
bioid=6,
prefix=,
role=,
orcid=0000-0002-2977-8911]
\ead{s.brinkkemper@uu.nl} 
%
\credit{Conceptualization, writing -- review \& editing}

\begin{abstract}
Software qualities such as \textit{usability} or \textit{reliability} are among the strongest determinants of mobile app user satisfaction and constitute a significant portion of online user feedback on software products, making it a valuable source of quality-related feedback to guide the development process. The abundance of online user feedback warrants the automated identification of quality characteristics, but the online user feedback's heterogeneity and the lack of appropriate training corpora limit the applicability of supervised machine learning. We therefore investigate the viability of three approaches that could be effective in \textit{low-data} settings: language patterns (LPs) based on quality-related keywords, instructions for crowdsourced micro-tasks, and large language model (LLM) prompts. We determined the feasibility of each approach and then compared their accuracy. For the complex multiclass classification of quality characteristics, the LP-based approach achieved a varied precision (0.38--0.92) depending on the quality characteristic, and low recall; crowdsourcing achieved the best average accuracy in two consecutive phases (0.63, 0.72), which could be matched by the best-performing LLM condition (0.66) and a prediction based on the LLMs' majority vote (0.68). Our findings show that in this low-data setting, the two approaches that use crowdsourcing or LLMs instead of involving experts achieve accurate classifications, while the LP-based approach has only limited potential. The promise of crowdsourcing and LLMs in this context might even extend to building training corpora. 
\end{abstract}

\begin{highlights}
\item Crowdsourcing and LLMs can support CrowdRE in eliciting quality requirements
\item Insufficient training data complicates automated quality classification in user feedback
\item Laypeople can accurately predict quality aspects through a series of crowdsourced micro-tasks
\item LLMs can under the right circumstances equate crowdsourced classification
\item Crowdsourcing and LLMs outperformed a linguistic approach in predicting qualities
\end{highlights}


\maketitle

\section{Introduction} \label{sec:introduction}
Engaging end-users and capturing their requirements are crucial for software system's success~\citep{Bano17}. A popular source of information is online user feedback~\citep{Astegher23}, which can be analyzed for user statements that express or pertain to requirements~\citep{Dabrowski22}. Approaches to (semi)automatically analyzing online user feedback in requirements engineering (RE) are commonly referred to as CrowdRE~\citep{Groen17ieee}. 

In addition to many CrowdRE works on functional requirements~\citep{Khan19}, some researchers have focused on identifying quality requirements~\citep[e.g.,][]{Groen17users,Jha17,Lu17}. Online user feedback addressing software product qualities can inform about \textit{how well} software delivers its functions, i.e., a system's \textit{qualities}. In particular, online user feedback has been found to address quality characteristics far more than functional aspects~\citep{Groen17users}. Especially negative experiences and opinions\textemdash such as poor usability or instability\textemdash have an impact on user satisfaction~\citep{Ceaparu04,Groen17users,Hertzum23}.

However, any approach aimed at classifying online user feedback\textemdash from manual to fully automated\textemdash faces difficulties due to informal language and finer nuances (e.g., Internet slang, expressions, sarcasm, emojis), author characteristics (e.g., laypeople, non-native English speakers and multilinguality, fakes), and general poor writing~\citep[e.g., sloppy use of punctuation, poor spelling, abundance of typos;][]{Groen18,Williams17}. Moreover, users provide a personal account in prose, the relevant software quality that causes (dis)satisfaction is often implicit, and users are not always able to describe problems well. 

\paragraph{Research problem.} 
Classifying online user feedback along quality dimensions faces unique challenges. This is due to not only the ambiguity of the data to be classified but also the way that taxonomies model quality dimensions. In computer science, the ISO/IEC 25010 standard~\citep{ISO11}, is the de facto standard with respect to software product qualities. It provides a taxonomy of eight characteristics\textemdash \textit{functional suitability}, \textit{performance efficiency}, \textit{compatibility}, \textit{usability}, \textit{reliability}, \textit{security}, \textit{maintainability}, and \textit{portability}\textemdash each with two or more subcharacteristics.\footnote{The experimental data on which this work reports was constructed when ISO/IEC 25010:2011 was in effect. By now, it has been superseded by the} ISO/IEC 25010:2023 standard~\citep{ISO23}, which introduces \textit{safety} as a ninth quality characteristic, renames \textit{usability} to \textit{interaction capability}, and \textit{portability} to \textit{flexibility}. Quality characteristics of software are typically specified in the form of \textit{quality requirements},\footnote{\citet{Glinz07} has suggested \textit{non-functional requirements} (NFRs) to be the overarching term for quality requirements and constraints. Our work does not address constraints, so we will speak of quality requirements where appropriate.} which are crucial for the success of software products. In RE, the ISO 25010 taxonomy is commonly used to organize quality requirements in a specification~\citep{Pohl15,Glinz23}, which is why we base our work on this particular taxonomy. However, because laypeople writing online user feedback often do not describe the circumstances in sufficient detail, it can be difficult to differentiate between these quality characteristics. For example, data getting corrupted might be primarily considered as \textit{integrity} (a subcharacteristic of \textit{security}), but if this happened during or because of an update, it could also be considered \textit{replaceability} (a subcharacteristic of \textit{portability}). The more ambiguous the feedback, the harder it is to correctly identify the affected quality. For instance, a statement such as ``app doesn't work'' could imply a problem with \textit{installability}, \textit{reliability}, \textit{compatibility}, \textit{interoperability} or \textit{security}, or maybe even reflect the user failing to understand the way the app is operated.

Although manual annotation by domain experts is the most accurate way of performing the difficult task of identifying software qualities in online user feedback, it is cognitively strenuous, time-consuming, and costly. Especially as the number of user reviews grows, automated analysis becomes indispensable~\citep{Groen18}. Finding a reliable alternative to expert-based annotations is an ongoing concern in CrowdRE. There are only a limited number of smaller annotated datasets~\citep[see][for a review]{Dabrowski22,Reddy21}, and these were created based on separate tagging schemata. This poses limitations to the use of machine learning (ML) classifiers, whose performance degrades when applied to unseen data, due to the heterogeneity of RE data~\citep{DellAnna23}. Hence, we call this domain a \textit{low-data setting}, referring to the absence of large and reliably labeled training corpora that can be used for training traditional ML and DL algorithms. In this setting, also based on initial experimentation, we decided to explore alternatives that do not require training data.

\paragraph{Solution approach.} 
To address the research challenge, we investigated the viability of approaches that do not require substantial training data. Specifically, we test each approach on the task of classifying online user reviews from app stores. We have selected three heterogeneous low-data approaches for this purpose:
\begin{itemize}
    \item \textit{Language patterns (LPs).} 
    Inspired by related work on linguistic approaches (see Section~\ref{sec:rw-nlp}), we adapted a method that was designed for eliciting non-functional requirements (NFRs), called the \textit{NFR Method}~\citep{Doerr11}, and constructed an approach for classifying software qualities based on predefined (combinations of) keywords without machine learning. We elicited taxonomies of keywords \& phrases through structured expert workshops, and then queried these over the online user feedback through LPs based on regular expressions. We considered this approach as a baseline against which the other approaches are tested, which is important to draw meaningful inferences.
    \item \textit{Crowdsourcing.} 
    We extended our previous research on the \textit{Ky\={o}ryoku} method, presented in~\citet{Vliet20}, to assess how accurately a crowd of paid laypeople can classify online user feedback. The Ky\={o}ryoku method was specifically designed to cater to crowd workers without prior knowledge of RE, who are, among other things, not familiar with the distinction between features and the qualities of these features. The crowd workers are given a brief training, and subsequently perform a micro-task. The instructions include examples, in line with the concept of few-shot learning in ML.
    \item \textit{Large language models (LLMs)}. 
    We developed a pipeline to prompt an LLM with context information and instructions to perform a classification task. Using this approach, we investigate how accurately different conditions based on LLM, prompt type, and learning strategy can classify online user feedback. 
\end{itemize}
\noindent Our categories of software product \textit{qualities} were based on ISO 25010. In some cases, they were renamed in accordance with the taxonomy of~\citet{Glinz07} to reduce task complexity and maximize comprehensibility for laypeople~\citep[cf.][]{Vliet20}. To help crowd workers understand better how \textit{reliability} is distinct from \textit{performance}, we chose \textit{stability} that reflects the most frequently addressed aspect of \textit{reliability} in online user feedback~\citep{Groen17users}. Because sentences in online user feedback often describe whether or not software supports a particular device or platform, it is difficult to determine whether this pertains to \textit{compatibility} or \textit{portability}, so that they were combined into a single class. Two ISO 25010 characteristics are not explicitly included: \textit{functional suitability} is difficult to comprehend and is implicitly covered by \textit{feature request}, and \textit{maintainability} is not visible to the user, but indirectly affects other qualities~\citep{Groen17users}. The category \textit{feature request} is included, so that the two main requirements types\textemdash functional and non-functional\textemdash are considered in the classification~\citep[cf.][]{Glinz07}.

\begin{table}[pos=h]
    \centering
    \caption{Overview of the ISO 25010 quality characteristics and their mapping onto the classes used for answering RQ2 and RQ3.}
    \label{tab:classes}
    \begin{tabular}{lll} 
		\hline
		\rowcolor[HTML]{D0D0D0} 
		\textbf{ISO 25010 Quality Characteristic} & \textbf{Ky\={o}ryoku Quality Aspect} & \textbf{Short Name} \\ \hline
		Compatibility \& Portability & System Support Feedback          & Compatibility                        \\ \hline
		Usability                    & User-Friendliness Feedback       & User-friendliness                    \\ \hline
		Security                     & Security Feedback                & Security                             \\ \hline
		Performance Efficiency       & Performance Feedback             & Performance                          \\ \hline
		Reliability                  & Stability Feedback               & Stability                            \\ \hline
		--                           & Feature Request                  & Feature                              \\ \hline
		--                           & None of the Above / Other        & None                                 \\ \hline
    \end{tabular}
\end{table}

\paragraph{Main research question.} 
We investigated each of the three approaches in an in-lab evaluation study, in which we seek to achieve an optimal configuration for each to answer this work's main research question (MRQ):
\vspace{5pt}
    \begin{enumerate}[\bfseries MRQ.]
        \item \textbf{Which low-data approach is most accurate in identifying quality aspects in online user feedback?}
    \end{enumerate}

\paragraph{Contributions.} 
The goal of this paper is to empirically investigate the effectiveness of and trade-offs between three low-data classification approaches. In doing so, this paper makes the following contributions:
\begin{itemize}
    \item We derive a keyword meta-model to help identify statements about quality aspects in online user feedback.
    \item We extend the crowdsourced user feedback classification method \textit{Ky\={o}ryoku}~\citep{Vliet20} with classifications into further quality aspects.
    \item We provide a state-of-the-art LLM pipeline to classify online user feedback according to requirements relevance and into particular qualities.
    \item We compare the three approaches that were fine-tuned to identify quality-related information in online user feedback.
\end{itemize}

\paragraph{Paper outline.} 
We begin by outlining our research methodology in Section~\ref{sec:method}. We then present the results for each research question in Section~\ref{sec:results}, and discuss our findings in Section~\ref{sec:discussion}. In Section~\ref{sec:ttv}, we present a discussion of the main threats to validity, Section~\ref{sec:rw} provides an overview of the relevant literature, and in Section~\ref{sec:conclusion}, we conclude. An online appendix provides supplementary material including additional content, experimental artifacts, and spreadsheets with data and results~\citep{appendix}.

\section{Research Methodology} \label{sec:method}
In this section, we first detail the research questions (RQs) that decompose our MRQ and the methods used (Section~\ref{sec:method_rqs}); then we introduce the dataset and gold standards (Section~\ref{sec:method_data}), and we describe the metrics and data analysis techniques we applied (Section~\ref{sec:method_metrics}). After that, we describe the experimental configuration for each of our four RQs in Sections~\ref{sec:method-kp}--\ref{sec:method-comp}.

\subsection{Research Questions \& Methods} \label{sec:method_rqs}
\begin{figure}[pos=h]
    \centering
    \includegraphics[width=.6\linewidth]{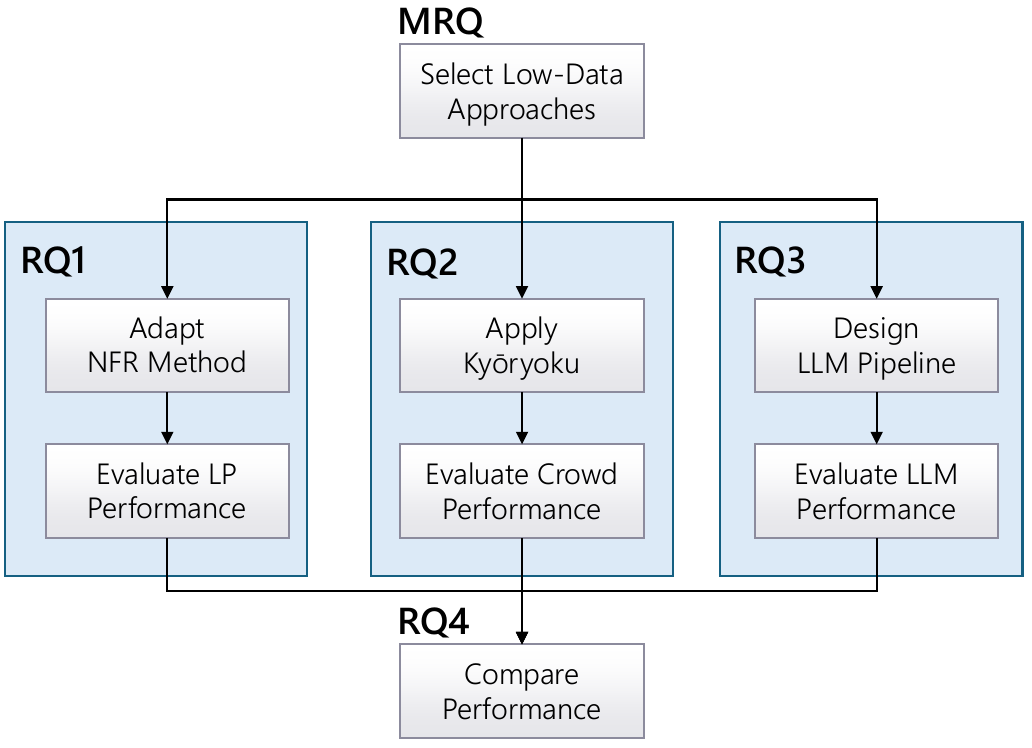}
    \caption{Illustration of the relationship between the main research question (MRQ) and the research questions (RQs).}
    \label{fig:RQs}
\end{figure}

In our investigation of which low-data approach is most accurate in identifying quality aspects in online user feedback, we have selected and developed three candidate approaches based on the literature. To achieve this, for each approach we followed the design science approach as suggested by~\citet{Wieringa14}, who distinguishes between two activities: a \textit{design activity} in which an artifact or other object is improved upon to address a practical problem, and an \textit{investigative activity} to investigate and scientifically evaluate this artifact in context. Our comparison takes into account that each of the chosen approaches has particular benefits and opportunities for research or practice, but also comes with its unique challenges and drawbacks. Therefore, we decomposed our MRQ into four RQs, which address the design--investigate cycles for each approach individually (RQ1--RQ3) before comparing them (RQ4), as visualized in Figure~\ref{fig:RQs}. 

    \begin{enumerate}[\bfseries RQ1.]
        \item \textbf{How to adapt the NFR Method in order to derive LPs for the identification of quality characteristics in online user feedback?}
    \end{enumerate}

\noindent We adapted the \textit{NFR Method} to derive LPs for the identification of quality characteristics in online user feedback, which is an empirical approach that aims to ensure that all quality requirements are elicited and correctly specified~\citep{Doerr11}. This is a difficult challenge in industry~\citep{Chung09}, which it addresses by incorporating concepts of various NFR approaches, including \textit{Goal-Oriented RE}~\citep{Lamsweerde01} and the \textit{NFR Framework}~\citep{Chung12}. The process adapts the backlog of (reusable) experience-based artifacts such as high-level quality attributes, custom quality models, templates, and checklists to a specific business and domain context. The process consists of two phases, each consisting of several activities. In Phase 1, the elicitation and specification are prepared by prioritizing the quality characteristics, preparing models, identifying potential conflicts between qualities, and deriving checklists and templates. In Phase 2, the quality requirements are elicited and dependencies between them are identified. Our online appendix~\citep{appendix} presents a detailed description of the NFR Method and its process.

In our work, we operationalized the NFR Method's strategy for the context of online user feedback about mobile apps. We first elicited relevant textual building blocks denoted \textit{keywords and phrases} (K\&Ps) and then refined them into \textit{language patterns} (LPs), which we used as queries to identify associated statements in online user feedback. In order to assess the merit of the designed approach based on LPs, we introduce and address research sub-question \textbf{RQ1a:}\textit{ How effective are language patterns based on quality-related keywords in identifying quality characteristics? }

\vspace{1em}

    \begin{enumerate}[\bfseries RQ2.]
        \item \textbf{How to design micro-tasks so that lay workers through crowdsourcing can achieve good results classifying quality aspects in online user feedback?}
    \end{enumerate}

\noindent To investigate the potential of lay workers to classify quality aspects in online user feedback, we further researched the Ky\={o}ryoku\footnote{Ky\={o}ryoku (\begin{CJK}{UTF8}{min}協力\end{CJK}) is a Japanese term for \textit{collaboration}: literally, it combines \textit{strength} (\begin{CJK}{UTF8}{min}力\end{CJK}) with \textit{cooperation} (\begin{CJK}{UTF8}{min}協\end{CJK}).} method introduced in~\citet{Vliet20}, which describes a crowdsourced annotation process to identify requirements-related content in online user reviews about software apps. Following \textit{CrowdForge}, a general-purpose framework on crowd work~\citep{Kittur11}, Ky\={o}ryoku involves a stepwise classification process through a series of micro-tasks. In crowdsourced work, \textit{micro-tasks} are structured and simplified data extraction decision workflows that are performed by \textit{crowd workers} in return for relatively small rewards~\citep{Retelny14,Valentine17}. Because each micro-task is more granular than the preceding one, they become increasingly demanding and complex for crowd workers~\citep[cf.][]{Schenk11,Gilardi23}, but this can be mitigated by making sure the micro-tasks’ instructions have been formulated to be simple enough for laypeople without relevant prior knowledge to understand them~\citep{Kittur11}. The method results in an annotated dataset that can, for example, be used as a gold standard to train ML classifiers. Each micro-task in Ky\={o}ryoku begins with a \textit{job description}, followed by an \textit{eligibility test} that needs to be passed to proceed to the actual \textit{annotation task}.\footnote{The job descriptions, test questions and data sample are available in the online appendix of~\citet{Vliet20}; \href{https://doi.org/10.5281/zenodo.3754721}{doi:10.5281/zenodo.3754721}.} In order to investigate the performance of crowd workers in different configurations, we proposed two research sub-questions that allow us to evaluate two main design choices of our approach:
       \begin{enumerate}[\bfseries {RQ2}a.]
              \item How does decomposing a classification task into smaller, consecutive classification tasks affect the performance of crowd workers?\\
              \textit{This sub-question compares the performance of the crowd in a condition in which the crowd workers assigned all quality aspects with a condition consisting of two micro-tasks in sequence.}
              \item Can crowd workers correctly differentiate between all quality aspects?\\
              \textit{This sub-question analyzes whether differences exist in the ability to recognize the various quality aspects.}
        \end{enumerate}

\vspace{1em}

    \begin{enumerate}[\bfseries RQ3.]
        \item \textbf{How to design an LLM} pipeline so that it can achieve good results in classifying quality aspects in online user feedback with little training data?
    \end{enumerate}

\noindent To determine whether an LLM pipeline is capable of correctly classifying online user feedback into quality dimensions with little training data (RQ3), we performed an experiment with a series of classification tasks similar to RQ2, in which we varied the \textit{large language model} (LLM), the prompt type, and the learning strategy. 
Our research design motivates two research sub-questions for RQ3:
    \begin{enumerate}[\bfseries {RQ3}a.]
        \item How does increasing task complexity affect the performance of LLMs?\\
        \textit{This sub-question compares the performance of LLMs} in phases with increasing complexity: P1, P2, and P3${^\prime}$.
        \item How does the use of engineered prompts and examples affect the performance of LLMs?\\
        \textit{This sub-question assesses the impact of the experimental factors prompt type and learning strategy.}
    \end{enumerate}

\vspace{1em}

    \begin{enumerate}[\bfseries RQ4.]
        \item \textbf{Which low-data approach is most suited for classifying quality aspects in online user feedback, considering their effectiveness and trade-offs?}
    \end{enumerate}

\noindent We compare the three low-data classification approaches in terms of their trade-offs and effectiveness in classifying online user feedback into quality aspects, which helps us to determine which approach is most accurate. We break down RQ4 into two research sub-questions: 
        \begin{enumerate}[\bfseries {RQ4}a.]
            \item How suitable is each approach for determining requirements relevance? \\
            \textit{This sub-question compares the performance of the crowd to that of LLMs} in terms of filtering out online user feedback irrelevant to RE (P1 \& P2).
            \item How suitable is each approach for distinguishing between the quality aspects? \\
            \textit{This sub-question compares the performance of all three approaches on the task of assigning the correct quality characteristics (P3 \& P4).}
        \end{enumerate}

\subsection{Dataset \& Gold Standards} \label{sec:method_data}

In order to allow comparisons between the three approaches considered, we used a shared subset of the large dataset, first introduced in Sect. II.A in~\citet{Groen17users}. This dataset contains 132,194 online user reviews curated from six \textit{app categories}\textemdash the five categories found by~\citet{Pagano13} to attract the most reviews, plus the emerging category ``smart products''\textemdash with two \textit{apps} per category\textemdash one paid, one free\textemdash and from three international English-language \textit{app stores}\textemdash Apple App Store, Google Play, and Amazon.com\textemdash for a total of 36 sources. Our subset omits Amazon's defunct app store for replicability reasons, and ``smart products'' apps because these were found to substantially differ from the other categories~\citep{Groen17users}, retaining 122,899 user reviews from 20 sources. 

We originally developed our gold standards in~\citet{Vliet20}, for which we took a data sample of 1,000 reviews that was systematically stratified across apps and app stores. This number was chosen to fit the job size limit of the account type used on the platform that we used in our study to answer RQ2. We created three gold standards used for a multi-phase analysis; we took the dataset from~\citet{Vliet20}, but revised it to make sure that each classification was judged by three raters, for which we reconciled disagreements: 
\begin{itemize}
    \item The gold standard for Phase P1 distinguishes user reviews into helpful and useless ones from the perspective of the developer of an app.
    \item The gold standard for Phase P2 distinguishes helpful user reviews from P1 into relevant and useless sentences from the perspective of the developer of an app.
    \item The gold standard for Phases P3 \& P4 distinguishes helpful sentences from P2 into quality aspects, \textit{feature request}, and \textit{none}.
\end{itemize}
\noindent Table~\ref{tab:goldstandards} shows how three online user reviews perpetuated through the gold standards. The 1,000 reviews used for the gold standards and that served as our \textit{test set} were omitted from the dataset to prevent overfitting. All artifacts related to the data and gold standards are available from~\citet{Vliet20} and the online appendix to this paper~\citep{appendix}.

After identifying an inconsistency in the data sample of~\citet{Vliet20}, in which text was erroneously inserted into some items, we performed data sanitation by manually removing reviews or sentences that could have been classified differently by the crowd. This resulted in the omission of 23 reviews in P1 (0.02\%), 635 sentences in P2 (51.12\%), 112 in P3 and P3${^\prime}$ (16.40\%), and 247 (19.03\%) in P4. This appears to have only affected our ability to draw conclusions about the \textit{useless} classifications in P2~\citep[see our online appendix for details on the data sanitation process and tables comparing the data;][]{appendix}. 
\begin{table}[pos=h]
    \centering
    \caption{Examples of online user reviews in the gold standards, with classifications in square brackets. Items classified as \textit{useless} are not perpetuated to the next phase. The sentence splitter for Phase P2 split sentences based on punctuation. In Phase P2, sentences that rely on the context to be understood are judged \textit{useless}. In Phase P3, multiple classifications were only assigned if a sentence was interpreted to ascribe about equal value to two different aspects.}
    \label{tab:goldstandards}
    \resizebox{\textwidth}{!}{%
    \renewcommand{\arraystretch}{1.2}
    \setlength{\tabcolsep}{4pt}
    \begin{tabular}{>{\raggedright\arraybackslash}p{0.32\textwidth}
                        >{\raggedright\arraybackslash}p{0.42\textwidth}
                        >{\raggedright\arraybackslash}p{0.36\textwidth}}
        \hline
	\rowcolor[HTML]{D0D0D0} \textbf{Gold Standard P1} & \textbf{Gold Standard P2} & \textbf{Gold Standard P3 \& P4} \\
        \hline
        \multirow{3}{=}{Can't access settings or feedback What good is an application if you can't change the settings or send feedback through it? I have the same problem with office lens. Typical arrogant Microsoft. \textbf{[Helpful]}} & Can't access settings or feedback What good is an application if you can't change the settings or send feedback through it? \textbf{[Helpful]} & Can't access settings or feedback What good is an application if you can't change the settings or send feedback through it? \textbf{[Feature Request, User-friendliness]} \\
        & I have the same problem with office lens. \textbf{[Useless]} & -- \\
        & Typical arrogant Microsoft. \textbf{[Useless]} & -- \\
        \hline
        App addict. This is the one and only and best app ever! Can't live without it! \textbf{[Useless]} & -- & \\
        \hline
        \multirow{4}{=}{Crashes when I open it \& terrible lag I've used this app for over 2 yrs \& have loved it until now. Every time I try to reply to someone the app closes. Also, the lag is terrible. This has been the best app until lately. \textbf{[Helpful]}} & Crashes when I open it \& terrible lag I've used this app for over 2 yrs \& have loved it until now. \textbf{[Helpful]} & Crashes when I open it \& terrible lag I've used this app for over 2 yrs \& have loved it until now. \textbf{[Stability]} \\
        & Every time I try to reply to someone the app closes. \textbf{[Helpful]} & Every time I try to reply to someone the app closes. \textbf{[Stability]} \\
        & Also, the lag is terrible. \textbf{[Helpful]} & Also, the lag is terrible. \textbf{[Performance]} \\
        & This has been the best app until lately. \textbf{[Useless]} & -- \\ \hline
    \end{tabular}
    }%
\end{table}

\subsection{Metrics \& Data Analysis} \label{sec:method_metrics}
For each approach, we compared the classifications against the gold standards to determine the true positives (TP), true negatives (TN), false positives (FP) and false negatives (FN). For each phase in RQ2, we created a combined prediction of the crowd workers' judgments by majority vote, with a tie constituting a multi-label decision. Inspired by the recommendation of~\citet{Mizrahi24} to evaluate LLMs by averaging them across multiple prompts, we similarly calculated a prediction by majority vote for RQ3. Based on these decision scores, we gauged the crowd's performance using the three primary performance metrics for evaluating classifier performance on a classification problem~\cite[cf.][]{DellAnna23}: 
\[Accuracy = \frac{TP+TN}{TP+TN+FP+FN}\]
\[\mathit{Precision} = \frac{TP}{TP+FP}\]
\[Recall = \frac{TP}{TP+FN}\]
\noindent Of these, \textit{accuracy} describes the fraction of the entire space that is classified correctly by a classifier, whether as a true positive or as a true negative; \textit{precision} measures the degree to which only the relevant answers are found, and \textit{recall} measures the degree to which all the relevant answers are found~\citep{Berry21}. Because our work explores the effects of both precision and recall, we report these scores instead of $F_1$/$F_\beta$ harmonized means of these two metrics. To compare the accuracy of the classification against the gold standards, we calculated correctness scores, constructed confusion matrices to identify where the most errors in a particular condition were made, and which two classes were confused most often. For cases in which the gold standard of P3 \& P4 allowed for multiple correct responses, the correct predictions were counted. If the majority vote was tied, resulting in a multi-label output, this response was considered correct if at least one prediction matched the gold standard; if all predictions were incorrect, these were counted as separate penalties for each incorrect class. We provide the macro average for precision and recall, and the mean score representing the accuracy value. To visualize differences in classifier performance, we created receiver operating characteristic plots~\citep[ROC Plots;][]{Fawcett06}, which contrasts \textit{sensitivity}\textemdash which is synonymous with recall\textemdash with \textit{specificity}:
\[\mathit{Specificity} = \frac{TN}{TN+FP}\]
Although for statistically analyzing classifiers\textemdash usually ML algorithms\textemdash \citet{DellAnna23} suggest repeated-measures designs, we consider the configuration of the LLM conditions for RQ3 to be a between-subjects design, because we had not controlled for non-determinism by altering the configuration of hyper-parameters\textemdash which is not possible for ChatGPT\textemdash or by sampling multiple outputs, and the various experimental factors inherently intended the LLMs to act differently across conditions. 

Our data was not normally distributed; Mardia's test for multivariate normality~\citep{Mardia70} revealed significant skewness from symmetry, $b_{1,p} = 1290.01$ (P1), 1534.57 (P2), 180.29 (P3${^\prime}$), $p < .001$, and significant kurtosis implying shorter tails than expected, $b_{2,p} = -44.81$ (P1), -53.39 (P2), -39.64 (P3${^\prime}$), $p < .001$. In order to measure the main and interaction effects of the three factors (prompt type, learning strategy, and LLM) on classification accuracy for Phases P1, P2, and P3${^\prime}$ and to compare LLM accuracy by phase, we used \textit{binomial Generalized Linear Models}~\citep[GLM;][]{Hosmer13}, which extend traditional linear regression to binary data for which normality cannot be assumed, and allow the modeling of non-linear relationships to estimate the effect of our experimental factors. Because we are also interested in how likely it is that a change in a condition alters the outcome, we quantify the strength of the effects using the \textit{Odds Ratio} (OR), which exponentially transforms the GLM's regression coefficient $\beta$ through the logistic function $OR = e^\beta$, and which serves as an indicator for the multiplication factor by which a change in condition is expected to increase or decrease~\citep{Kutner05}. Because we found some results that would have been significant had the data been normalized, we also report results that show a trend toward significance. Because from early on, we had indication that the LLM used had the greatest impact on the results, we often contrasted the performance of ChatGPT 4 Legacy and ChatGPT 4o. 

To answer RQ4, we compared the performance of the three approaches in several ways. For comparing classifier performance, we contrasted the means with Tukey's Honestly Significant Difference (HSD) test for pairwise comparisons~\citep{Tukey49}. To compare the performance of the classifiers by class, we analyzed them individually as binary predictions. This did impact the accuracy of the classifiers based on crowdsourcing and ChatGPT, because it entailed assigning penalties for each missed multi-label options in the gold standard and each incorrectly provided multi-label prediction. Because a comparison of performance on individual classes constitutes a repeated-measures design, we compared the performance of the classifiers on individual classes using the Friedman Test~\citep{DellAnna23}, which for significant $\chi^2$ scores we further explored using Wilcoxon signed-rank tests for pairwise comparisons with Bonferroni-adjusted $p$-values~\citep[cf.][]{Demsar06}. We contrasted the classifier with the highest mean per class against the other scores using binomial GLMs. 

\subsection{Experimental Configuration for RQ1 -- Language Patterns} \label{sec:method-kp}
To answer RQ1, we elicited keywords and phrases (K\&Ps) through a series of workshops adapted from the NFR Method~\citep{Doerr11}, based on which we created language patterns (LPs). To systematically elicit K\&Ps for the quality characteristics of ISO 25010, we conducted six elicitation workshops of a maximum of 1.5 hours with a total of 24 participants (11 female, 13 male) solicited from among Computer Science students at two universities, doctoral candidates in Computer Science, and scientific staff at Fraunhofer IESE, a research institute. The workshop addressing the characteristics of \textit{usability}, \textit{security} and \textit{maintainability} was attended by the same experts. Most of the participants were German (11) or Dutch (7) and had good self-reported English proficiency. Despite the fact that all of the participants had been using a smartphone for many years, only five had ever written a user review. Due to the COVID-19 pandemic, the final two workshops were held remotely. Each workshop consisted of five steps: (1)~briefing and formalities, (2)~aligning on the quality characteristic, (3)~establishing the quality model, (4)~defining K\&Ps, and (5)~debriefing. To this end, the participants received the ISO 25010 definitions of their workshop's quality characteristic and its subcharacteristics, written guidelines that were designed to ensure lively interaction during the workshop, and templates to develop the understanding of the participants about the quality characteristic and for collecting the K\&Ps. 

After the workshops, we documented the outcomes and calculated descriptive statistics. We recoded the elicited K\&Ps into LPs using the regular expressions notation, so that the K\&Ps can be searched automatically through queries. Because the K\&P were obtained during brainstorming sessions in which participants had little time to reflect on the results, and because the participants looked at only one characteristic, two authors served as judges to realign the classification, which included the inclusion of synonyms, variations, negated antonyms, and forbidden words to prevent false positives (FPs). This resulted in 248 LPs, which we queried over our analysis set (Round 1). We analyzed all results for the LPs with up to 100 matched statements, or a stratified random sample of 100 matched statements for LPs with more matches, so that in total, we reviewed 7,467 statements. We changed the regular expression syntax to exclude FPs where possible to increase precision. An example of an LP for the \textit{portability} subcharacteristic \textit{replaceability} is: \texttt{(?i)(?<!should |could |would |th)(is |are |has |have )(a |an |)(far |much |)(more |)(improv|upgrade|faster|quicker)}. We repeated the query and the analysis over a sample of 6,149 statements (Round 2). LPs with a precision below $0.50$ were discarded. The final set of 242 LPs was used to query our test set to measure precision and recall against the gold standard for P3 \& P4. The online appendix~\citep{appendix} provides a detailed description of the relatively complex LP encoding process, and a primer on regular expression notation. 

\subsection{Experimental Configuration for RQ2 -- Crowdsourcing} \label{sec:method-cs}
To answer RQ2, we conducted a single-group experiment for which we recruited crowd workers through the online crowdsourcing marketplace \textit{Figure Eight},\footnote{Later acquired by Appen, \url{https://www.appen.com/}} which allows crowd workers to perform jobs by assigning micro-tasks in exchange for fixed-price monetary rewards. This section presents an abridged overview of the method presented in~\citet{Vliet20}, extended by Phases P4 and P3${^\prime}$.

\begin{figure}[pos=h]
    \centering
    \includegraphics[width=1\linewidth]{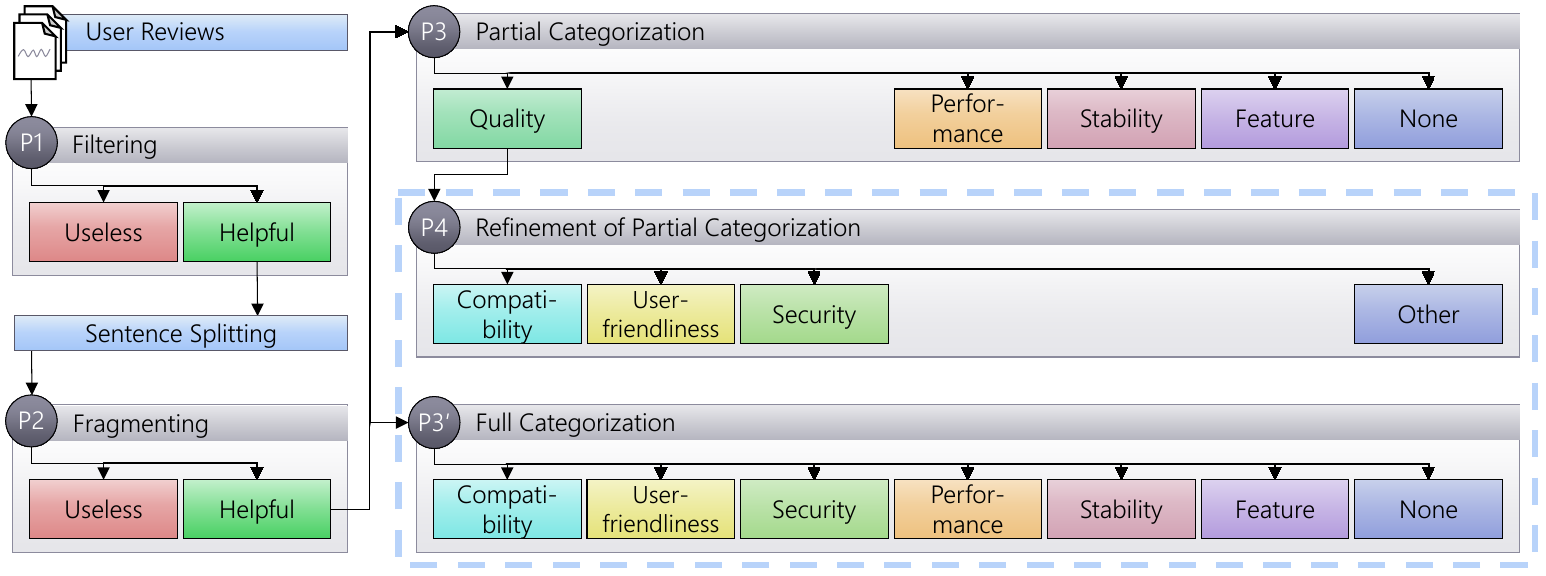}
    \caption{Overview of the classifications in the various phases performed using the Ky\={o}ryoku method. Our work extends~\citet{Vliet20} with two phases, P4 and P3${^\prime}$, shown using a dashed line.}
    \label{fig:phases}
\end{figure}

We selected \textit{Figure Eight} because of its support for data categorization tasks and its many embedded quality control mechanisms. Following the Ky\={o}ryoku method, a set of 1,000 user reviews was classified in consecutive micro-tasks, initially distinguishing between requirements-relevant and requirements-irrelevant~\citep[cf.][]{Sihag23}, and later between various quality characteristics. Figure~\ref{fig:phases} visualizes the sequence of the phases, which are as follows:
\begin{itemize}
  \item[\bfseries P1] The annotation task of P1 served to determine whether the body text of unprocessed user reviews could be \textit{helpful} to software developers or should be considered \textit{useless}.
  \item[\bfseries P2] The \textit{helpful} reviews from P1 were split into individual sentences that the crowd workers in P2 distinguished into \textit{helpful} and \textit{useless}.
  \item[\bfseries P3] The \textit{helpful} sentences obtained from P2 underwent a more fine-grained classification (see Table~\ref{tab:classes} in Section~\ref{sec:introduction}). In~\citet{Vliet20}, we reported on Phase 3 (\textbf{P3}), in which the crowd workers performed a partial classification in which the classes \textit{compatibility}, \textit{user-friendliness} and \textit{security} were combined into the class \textit{quality feedback}.
  \item[\bfseries P4] In this work, the \textit{quality feedback} class served as input for P4. The class \textit{none of the above} was used for aspects such as general criticism and praise; in P4 this category was called \textit{other}, because based on the classification of the preceding P3, it already described a quality of some sort.
  \item[\bfseries \textbf{P3${^\prime}$}] We furthermore investigated the crowd workers’ performance when classifying all classes at once in Phase 3 Primed (P3${^\prime}$), in which the three quality aspects combined in P3 as \textit{quality feedback} were presented together with the other quality aspects.
\end{itemize}

We refer to the first two phases, which focus on requirements-relevance, as \textbf{P1 \& P2}. Because P3 and P4 together constitute a classification of all quality aspects in two consecutive phases, we jointly refer to those as \textbf{P3$\rightarrow$P4}. We collectively refer to the combination of Phases P3$\rightarrow$P4 and P3${^\prime}$ as \textbf{P3 \& P4}. Table~\ref{tab:configuration} shows the configuration for these five phases, with P3 \& P4 enabling us to compare two variants: a single session with all classes (P3${^\prime}$) and two consecutive sessions (P3$\rightarrow$P4). In \textit{Figure Eight}, we opted for an open crowd selection policy; it is realistic to expect non-native English-speaking crowd workers to be capable of performing such a task, and their language proficiency could be ensured through the eligibility test. Three crowd worker judgments were sufficient to determine the majority vote for the binary classifications in P1 and P2. For the multiclass classifications in P3 \& P4, we based the majority vote on six crowd worker judgments. The total cost of our experiment was \$619.32, which includes \textit{Figure Eight}'s 20\% usage fee. The sessions were active for 10:44:47 hours in total before reaching completion.

\begin{table}[pos=h]
    \centering
    \caption{Summary of the configuration for each phases offered on \textit{Figure Eight}, along with runtime statistics per session. P1 and P2 consisted of two micro-tasks. The sequence representation illustrates how Phases P3$\rightarrow$P4 are performed concurrently to Phase P3${^\prime}$.}
    \label{tab:configuration}
    \resizebox{0.75\textwidth}{!}{%
    \begin{tabular}{l|l|l|c|r|c|S[table-format=4.0]|l|S[table-format=3.2]}
		\hline
		\rowcolor[HTML]{D0D0D0} 
		\textbf{Sequence} & \textbf{Phase} & \multicolumn{1}{c|}{\begin{tabular}[c]{@{}c@{}}\textbf{Launch}\\ \textbf{(2019)}\end{tabular}} & \textbf{\begin{tabular}[c]{@{}c@{}}Judg- \\ ments\end{tabular}}  & \textbf{Duration} & \textbf{\begin{tabular}[c]{@{}c@{}}Jugments\\ per Item \end{tabular}} & \textbf{\begin{tabular}[c]{@{}c@{}}Contri- \\ butors\end{tabular}} & \textbf{\begin{tabular}[c]{@{}c@{}}Price per\\ Judgment (\$)\end{tabular}} & \textbf{\begin{tabular}[c]{@{}c@{}}Total\\ Cost (\$)\end{tabular}} \\ \hline
		\multirow{5}{*}{
\begin{tikzpicture}
    \draw[line width=0.5mm,black,densely dotted] (0.1,1.9) -- (0.1,0.9); 
    \draw[line width=0.5mm,black,densely dotted] (1.1,1.3) -- (1.1,1.9); 
    \draw[line width=0.5mm,black,densely dotted] (0.1,1.9) -- (1.1,1.9); 
    \draw[line width=0.5mm,black,densely dotted] (0.6,2.5) -- (0.6,1.9); 

    \fill[darkgray] (0,0.8) rectangle (0.2,1.0); 
    \fill[darkgray] (1.0,1.2) rectangle (1.2,1.4); 
    \fill[darkgray] (1.0,1.6) rectangle (1.2,1.8); 
    \fill[darkgray] (0.5,2.0) rectangle (0.7,2.2); 
    \fill[darkgray] (0.5,2.4) rectangle (0.7,2.6); 
    
    \fill[white] (0,2.6) rectangle (0.6,2.67); 
\end{tikzpicture}
		} 
		 & P1            & May 7 \& 15  & 50 & 1:48:40 & 3 & 123 & ~~~0.03--0.04 & 130.80  \\
		 & P2            & May 23 \& 29 & 50 & 1:10:12 & 3 & 235 & ~~~0.02    & 106.32  \\
		 & P3            & June 13      & 50 & 2:19:21 & 6 & 145 & ~~~0.02    & 117.60  \\ 
		 & P4            & Dec. 27      & 50 & 1:58:36 & 6 & 62  & ~~~0.03    & 77.04   \\ 
		 & P3${^\prime}$ & Dec. 27      & 50 & 3:27:58 & 6 & 146 & ~~~0.03    & 187.56  \\ \hline
		~ & \textbf{Total} &  &  & \textbf{10:44:47} &  & \textbf{711} &  & \textbf{619.32} \\ \hline
    \end{tabular}%
    }
\end{table}

\subsection{Experimental Configuration for RQ3 -- LLMs} \label{sec:method-llms}
To answer RQ3, we performed a between-subjects experiment with a 2 (prompt type) $\times$ 2 (learning strategy) $\times$ 2 (LLM) factorial design, for a total of eight conditions, comparing the following experimental factors:

\begin{enumerate}
    \item Two \textit{prompt types}: engineered (\textit{Eng}) vs. Ky\={o}ryoku (\textit{Ky\={o}});
    \item Two \textit{learning strategies}: few-shot (\textit{Few}) vs. zero-shot (\textit{Zer});
    \item Two \textit{LLMs}: ChatGPT 4 Legacy\footnote{Previously known as ChatGPT 4 Turbo, OpenAI renamed it ChatGPT 4 Legacy upon the release of its successor model, ChatGPT 4o.} (\textit{4}) vs. ChatGPT 4o (\textit{4o}).
\end{enumerate}

\noindent Each condition was applied to the classification problems of Phases P1 (binary), P2 (binary) and P3${^\prime}$ (multiclass, multi-label) described in Section~\ref{sec:method-cs}. Due to the novelty of the domain, we compensated for the lack in established approaches with an elaborate pretest and by making improvements during the experiment. In the following, we briefly discuss the most important decisions, which are further elaborated in our online appendix~\citep{appendix}.

In a pretest, we compared Mixtral, Nous Hermes, ChatGPT 3.5 and the then-latest ChatGPT 4. We found that ChatGPT was the most capable of handling our prompts by their number of tokens and producing responses that were meaningful and correctly structured. Because this study serves to provide an estimation of the capability of LLMs rather than a comprehensive comparison of LLMs, we chose to perform this study using the state of the art in LLMs based on the literature available at the time~\cite[e.g.,][]{Gilardi23}. Our experiment was set up and trialed using ChatGPT 4 Legacy, which had been superseded by ChatGPT 4o by the time we conducted it. Because we had not studied the characteristics of the then just released ChatGPT 4o, we decided to compare the performance of both LLMs. A downside is that the GPT models hosted by OpenAI are black boxes of which the evolution cannot be controlled, impairing its replication potential compared to other LLMs. We did not fine-tune or retrain models due to the unavailability of training data in our low-data setting. ChatGPT on the other hand usually hallucinates less than other LLMs, even though its hyper-parameter settings cannot be altered from the web interface. ChatGPT was configured without custom instructions or memory function. A detailed discussion of what implications the choice of ChatGPT had on the validity of our work is provided in Section~\ref{sec:ttv}. 

Like~\citet{Gilardi23}, who provided the exact same instructions to ChatGPT that they had given their crowd workers, we use the instructions from the Ky\={o}ryoku method that we used for our micro-tasks and our gold standards. However, because there was also some evidence that a prompt specifically engineered for ChatGPT might lead to better results~\citep{Brand23, ElHajjami24}, we also engineered a prompt, which included the examples that we also provided to the crowd workers, making this a few-shot prompting strategy. As an example, Figure~\ref{fig:abridged_prompt} shows the few-shot engineered prompt for P1. The literature suggests that the behavior of ChatGPT can be influenced both positively and negatively by the use of examples, so we also included a learning strategy factor in which we compared the effect of the prompt including examples (i.e., a few-shot prompting strategy) with one in which these were omitted (i.e., a zero-shot prompting strategy). We presented the input data using text-based prompts in batches of 100 items or however many remained: i.e., 10 batches for P1 (1,000 reviews), 14 batches for P2 (1,347 sentences), and 7 batches for P3${^\prime}$ (625 sentences). The 31 batches across eight conditions amounted to a total of 248 treatments. To prevent contamination through learning effects, each treatment was performed in a new prompt session after restarting the Web browser. Because two responses per condition were found to often be sufficiently similar~\citet[e.g.,][]{Gilardi23}, we chose to collect one response per condition and instead focus on comparing other experimental factors that, to our knowledge, had not yet been empirically tested.

\begin{figure}[!htb]
\centering
\fbox{\begin{varwidth}{\dimexpr\textwidth-2\fboxsep-2\fboxrule\relax}
\footnotesize
\textbf{\# Binary Classification Task for ChatGPT}
\vspace*{0.4cm}

\textbf{\#\# Task Description}

Classify user reviews from mobile app stores (Google Play Store, Apple App Store) as either "helpful" or "useless". The objective is to filter out spam and irrelevant reviews, aiding developers in focusing on constructive feedback.

\vspace*{0.4cm}
\textbf{\#\# Context}

Developers depend on user reviews for actionable feedback. The task aims to pre-process these reviews, filtering out useless reviews that are irrelevant, whilst retaining helpful reviews.

\vspace*{0.4cm}
\textbf{\#\# Input Data Source}

The data to classify is provided at the end of this prompt under `Reviews to Classify'.

\vspace*{0.4cm}
\textbf{\#\# Class Definitions}

- **Useless**: Reviews that are not helpful for developers. They may contain spam, be off-topic, express feelings without elaboration, lack genuineness, or contain jokes.

- **Helpful**: Reviews that offer specific feedback, report bugs, suggest improvements, comment on updates, or provide constructive criticism.

\vspace*{0.4cm}
\textbf{\#\# Examples for Each Class}

\vspace*{0.2cm}
**Class 1: Useless**

- Example: ``I Really Like This!''

- Example: ``My kids love it. Thanks''

\vspace*{0.2cm}
**Class 2: Helpful**

- Example: ``Newest version crashes when opening''

- Example: ``Buggy and unreliable. Does not work often. Signs me out regularly.''

\vspace*{0.2cm}
Note: The examples provided are real user reviews. They may not always follow conventional punctuation or stylistic norms, reflecting the authentic and varied nature of user-generated content.

\vspace*{0.4cm}
\textbf{\#\# Instructions for Classification}

1. Assess each review in the input data.

2. Classify each review as either **Helpful** or **Useless** based on the definitions and examples provided.

\vspace*{0.4cm}
\textbf{\#\# Output Format}

Present results in a numbered list providing three values on a single line:

  - **Review**: The first three words from the input data, for reference.

  - **Judgment**: Your assigned classification as Useless or Helpful.

  - **Confidence**: A qualitative assessment of your confidence in the classification as High, Medium, or Low.

Provide only the values specified. Do not provide a justification for your judgment.

\vspace*{0.4cm}
\textbf{\#\# Additional Considerations}

- Begin the classification process by directly addressing all reviews below. Avoid trial runs or partial classifications; instead, apply the classification criteria to the entire dataset in a single, comprehensive review process.

- Strictly adhere to the provided class definitions when classifying reviews. Ensure that the classification as either `useless' or `helpful' is firmly based on the criteria specified, without leniency. This is crucial for maintaining the integrity and accuracy of the classification process.

- Consider each review in isolation. When classifying, if a review's content is unclear, ambiguous, or incomprehensible, use your best judgment. Classify a review as `Useless' if its content is so unclear, ambiguous, or incomprehensible that it would not reasonably make sense to a human reader.

- Approach the classification with the understanding that examples serve as guidelines but may not cover all scenarios. Use judgment to classify reviews that may not fit neatly into one category.

- In cases where a review contains both helpful and useless elements, classify it as `Helpful'. This approach ensures that potentially valuable feedback is not overlooked.

\vspace*{0.4cm}
\textbf{\#\# Reviews to Classify}

\ldots
\end{varwidth}}
\caption{Abridged engineered prompt for Phase 1 with examples, i.e., $\lcurvyangle$Eng,\,Few$\rcurvyangle$. See Figure 2 in~\citet{Vliet20} for a comparison to the Ky\={o}ryoku prompt, i.e., $\lcurvyangle$Ky\={o},\,Few$\rcurvyangle$. We omitted some examples for brevity. See the online appendix~\citep{appendix} for the full set of prompts.}
\label{fig:abridged_prompt}
\end{figure}

\subsection{Experimental Configuration for RQ4 -- Comparison} \label{sec:method-comp}
To compare the performance of the three approaches (RQ4), we conducted a posttest-only nonequivalent groups quasi-experiment~\citep{Jhangiani19} to obtain a performance benchmark between nonequivalent groups, with the classification problem as the independent variable. To predict requirements relevance in Phases P1 and P2, we compared the crowd's majority vote prediction (RQ2) against the best-performing ChatGPT condition and the majority vote prediction of the eight ChatGPT conditions (RQ3). Our LP approach (RQ3) did not predict requirements-relevance. Because ChatGPT had an even number of conditions, 104 predictions in P1 and 99 predictions in P2 of these binary classification tasks resulted in a 4:4 tie, which we omitted from the analysis to prevent undue penalties. We compared the performance of all three approaches over the gold standard for P3 \& P4 for five conditions:
\begin{itemize}
    \item \textit{Language Patterns.} The prediction through the LP-based analysis through 242 LPs obtained using the NFR Method.
    \item \textit{Crowd P3$\rightarrow$P4.} The majority vote prediction of six crowd worker judgments, respectively, in Phase P3$\rightarrow$P4 of the crowdsourced approach using Ky\={o}ryoku.
    \item \textit{Crowd P3${^\prime}$.} The majority vote prediction of six crowd worker judgments in Phase P3${^\prime}$ of the crowdsourced approach using Ky\={o}ryoku.
    \item \textit{ChatGPT Best.} The best-performing ChatGPT condition $\lcurvyangle$Ky\={o},\,Zer,\,4o$\rcurvyangle$ in P3${^\prime}$ by absolute positives in the LLM-based approach.
    \item \textit{ChatGPT Majority Vote.} The majority vote prediction of the eight ChatGPT conditions in P3${^\prime}$ in the LLM-based approach.
\end{itemize}

\section{Results} \label{sec:results}
In this section, we present the results towards answering our research questions for each of the four research questions.

\subsection{Results for RQ1 -- Language Patterns} \label{sec:results-kp}
Table~\ref{tab:results_short} summarizes the quantitative results of the workshop outcomes and the LPs encoded from the K\&Ps, which are further detailed in the online appendix~\citep{appendix}. The workshops produced a good number of 234 K\&Ps, ranging from 23 for \textit{portability} to 54 for \textit{performance efficiency}. After encoding the 248 LPs, they matched 143,044 statements in Round 1, and achieved an overall weighted micro-average precision of 65\%, meaning that two-thirds of the statements matched by all LPs were true positives (TPs). After making adjustments, the 242 LPs in Round 2 matched 77,935 statements ($\mbox{-}45.5\%$). The precision improved considerably for all characteristics, reaching a micro-average of $0.87$ overall. In most cases, a minor change could increase a particular LP's precision in Round 2, and like~\citet{ClelandHuang07}, we found that certain keywords were weak indicators by themselves (e.g., ``available'' for \textit{availability}). Almost all LPs improved; 115 of them even reached perfect precision. The five LPs from Round 2 that still had a precision below $0.5$ were removed, leading to a final set of 237 LPs. 

\begin{table}[pos=h]
    \centering
    \caption{Number of K\&Ps by ISO 25010 quality characteristic, and number of LPs (N) and precision (P) achieved per round. Red and green colors indicate the highest and lowest precision value by class.}
    \label{tab:results_short}
    \setlength{\tabcolsep}{5pt}
    \scriptsize
    \begin{tabular}{l|S[table-format=3.0]|S[table-format=3.0]S[table-format=1.2]|S[table-format=3.0]S[table-format=1.2]} \hline
         \rowcolor[HTML]{D0D0D0} 
         & \textbf{Elicited} & \multicolumn{2}{c|}{\textbf{LPs Round 1}} & \multicolumn{2}{c}{\textbf{LPs Round 2}} \\
         \rowcolor[HTML]{D0D0D0} 
         & \textbf{K\&Ps} & \multicolumn{2}{c|}{\textbf{(N, P)}} & \multicolumn{2}{c}{\textbf{(N, P)}} \\ \hline
         Compatibility    & 42 & 41 & \cellcolor[HTML]{E0FFE0} 0.75 & 41 & 0.91 \\
         Portability      & 23 & 52 & 0.59 & 27 & 0.88 \\
         Usability        & 40 & 36 & 0.57 & 24 & \cellcolor[HTML]{FFE0E0} 0.78 \\
         Security         & 51 & 31 & \cellcolor[HTML]{FFE0E0} 0.54 & 20 & \cellcolor[HTML]{E0FFE0} 0.92 \\
         Perf. Efficiency & 54 & 57 & 0.74 & 37 & 0.88 \\
         Reliability      & 24 & 31 & 0.69 & 21 & \cellcolor[HTML]{E0FFE0} 0.92 \\ \hline
         \textbf{Total} & \textbf{234} & \textbf{248} & \textbf{0.65} & \textbf{242} & \textbf{0.87} \\ \hline
    \end{tabular}
\end{table}

Table~\ref{tab:confusionkp} shows the confusion matrix for the performance of the LPs compared to the gold standard for P3 \& P4. Compared to Round 2, the LPs had a lower but still fair precision in the 0.40--0.97 range, but achieved low recall in the 0.06--0.50 range. Because the pattern matching only assigned positives (i.e., statements it positively matched), we considered all missed classifications as \textit{none}, resulting in limited precision (0.31) but high recall (0.95) on this class because it missed many statements but also matched some statements that in the gold standard are labeled as \textit{none}. \textit{Reliability} achieved good results, with a precision of 0.97 and a recall of 0.50. The combination of \textit{interoperability} and \textit{portability} achieved the poorest results, with a precision of 0.40 and a recall of 0.06. 

\begin{table}[pos=h]
    \centering
    \caption{Confusion matrix comparing the LP results to the gold standard for P3 \& P4. All aspects not matched by an LP were labeled \textit{none / missed}.}
    \label{tab:confusionkp}
    \scriptsize
    \begin{tabular}{l|S[table-format=2.0]S[table-format=3.0]S[table-format=2.0]S[table-format=2.0]S[table-format=2.0]S[table-format=3.0]|S[table-format=1.0]S[table-format=2.0]|S[table-format=1.2]S[table-format=1.2]}
		\hline
		\rowcolor[HTML]{D0D0D0} \cellcolor[HTML]{E0E0E0} & \multicolumn{8}{c|}{\textbf{Gold Standard}} & & \\
		\rowcolor[HTML]{D0D0D0} 
		\cellcolor[HTML]{E0E0E0} \textbf{LPs} & \textit{\begin{tabular}[c]{@{}c@{}}Compati- \\ bility\end{tabular}} & \textit{\begin{tabular}[c]{@{}c@{}}User- \\ Friendl.\end{tabular}} & \textit{\begin{tabular}[c]{@{}c@{}}Secu- \\ rity\end{tabular}} & \textit{\begin{tabular}[c]{@{}c@{}}Perfor- \\ mance\end{tabular}} & \textit{\begin{tabular}[c]{@{}c@{}}Stabi- \\ lity\end{tabular}} & \textit{\begin{tabular}[c]{@{}c@{}}None / \\ Missed\end{tabular}} & \textit{\begin{tabular}[c]{@{}c@{}}Mult. \\ (Corr.)\end{tabular}} & \textit{\begin{tabular}[c]{@{}c@{}}Mult. \\ (False)\end{tabular}} & \textbf{\begin{tabular}[c]{@{}c@{}}Pre- \\ cision\end{tabular}} & \textbf{\begin{tabular}[c]{@{}c@{}}Re- \\ call\end{tabular}} \\ \cline{2-11}
		\cellcolor[HTML]{E0E0E0} \textit{Interop./Port.} & \cellcolor[HTML]{F0F0F0} 4  & 2   & 0  & 0  & 1  & 0   & 0 & 3  & 0.40 & \cellcolor[HTML]{FFE0E0} 0.06 \\
		\cellcolor[HTML]{E0E0E0} \textit{Usability}      & 1  & \cellcolor[HTML]{F0F0F0} 11  & 2  & 0  & 0  & 0   & 1 & 0  & 0.81 & 0.10 \\
		\cellcolor[HTML]{E0E0E0} \textit{Security}       & 2  & 4   & \cellcolor[HTML]{F0F0F0} 4  & 0  & 1  & 0   & 1 & 0  & 0.42 & 0.24 \\
		\cellcolor[HTML]{E0E0E0} \textit{Performance}    & 2  & 1   & 0  & \cellcolor[HTML]{F0F0F0} 4  & 0  & 5   & 1 & 0  & 0.47 & 0.13 \\
		\cellcolor[HTML]{E0E0E0} \textit{Reliability}    & 0  & 1   & 0  & 0  & \cellcolor[HTML]{F0F0F0} 50 & 1   & 5 & 0  & \cellcolor[HTML]{E0FFE0} 0.97 & 0.50 \\
		\cellcolor[HTML]{E0E0E0} \textit{None / Missed}  & 62 & 112 & 11 & 42 & 50 & \cellcolor[HTML]{F0F0F0} 140 & 0 & 28 & \cellcolor[HTML]{FFE0E0} 0.31 & \cellcolor[HTML]{E0FFE0} 0.95 \\ \hline
		\cellcolor[HTML]{E0E0E0} \textit{Mult. (Corr.)}  & 0  & 1   & 0  & 2  & 1  & 0   & 3 & ~  &      &      \\
		\cellcolor[HTML]{E0E0E0} \textit{Mult. (False)}  & 0  & 0   & 0  & 0  & 0  & 1   & ~ & 0  &      &      \\ \hline
		\cellcolor[HTML]{D0D0D0} \textbf{Macro Average}  & \multicolumn{8}{c|}{} & 0.56 & 0.33 \\
		\cellcolor[HTML]{D0D0D0} \textbf{Accuracy}   & \multicolumn{8}{c|}{} & \multicolumn{2}{c}{0.41} \\ \hline
    \end{tabular}
\end{table}

\subsection{Results for RQ2 -- Crowdsourcing} \label{sec:results-cs}
We first describe the crowd that we assembled through \textit{Figure Eight} and the job they performed (Section~\ref{sec:demo}), and then we report on the accuracy of the crowdsourced work in terms of precision and recall by classification output (Section~\ref{sec:classacc}) and agreement (Section~\ref{sec:agreeacc}).

\subsubsection{Crowd Demographics \& Job Statistics} \label{sec:demo}
We gathered a large worldwide crowd through multiple crowd work channels associated with \textit{Figure Eight}. A total of 711 unique crowd workers commenced participation in one of the seven micro-tasks, 555 (78.06\%) of whom contributed judgments beyond the eligibility test and passed the quality checks. These 555 workers can be considered contributors. Most of the crowd workers were from countries where parts of the population live in poverty, such as Venezuela~\citep{Posch22}, which alone accounted for 33.47\% of all contributors.

We received a total of 16,428 annotations (24,307 including trial data), with an average of 35.72 classified items per contributor. This data includes trial data and actual judgments. A contributor who completes all 50 judgments generates 14 trial entries; ten from the first page acting as an aptitude test, while on the next four pages, one in ten items was a test question per page.  Table~\ref{tab:pages} shows that on the whole, half of the participants who started a session completed all five pages.

\begin{table}[pos=h]
    \centering
    \caption{Number of pages completed by each contributor per session. That two participants were able to complete six pages in Phase P3 is presumably due to an anomaly in \textit{Figure Eight}.}
    \label{tab:pages}
    \begin{tabular}{l|S[table-format=3.0]S[table-format=3.2]|S[table-format=3.0]S[table-format=3.0]S[table-format=3.0]S[table-format=2.0]S[table-format=3.0]}
		\hline
		\rowcolor[HTML]{D0D0D0} 
		\textbf{\begin{tabular}[c]{@{}c@{}}No. of \\ Pages\end{tabular}} & \multicolumn{2}{c|}{\textbf{\begin{tabular}[c]{@{}c@{}}Participants \\ (N,\,\%)\end{tabular}}} & \textbf{P1} & \textbf{P2} & \textbf{P3}  & \textbf{P4} & \textbf{P3${^\prime}$}  \\ \hline
		<1    & 21  & 2.95   & 2  & 11 & 1   & 2  & 5   \\ 
		1     & 150 & 21.10  & 16 & 89 & 24  & 9  & 12  \\ 
		2     & 71  & 9.99   & 14 & 28 & 5   & 10 & 14  \\ 
		3     & 45  & 6.33   & 11 & 17 & 4   & 6  & 7   \\ 
		4     & 62  & 8.72   & 27 & 15 & 8   & 1  & 11  \\ 
		5     & 360 & 50.63  & 53 & 75 & 101 & 34 & 97  \\ 
		6     & 2   & 0.28   &    &    & 2   &    &     \\ \hline
		\textbf{Total} & \textbf{711} & \textbf{100.00} & \textbf{123} & \textbf{235} & \textbf{145} & \textbf{62} & \textbf{146} \\
		\hline
    \end{tabular}
\end{table}

Our experimental setting resulted in the two scenarios illustrated earlier in Table~\ref{tab:configuration}. For the sequence P1$\rightarrow$P2$\rightarrow$P3${^\prime}$, we attracted 504 crowd workers over a time span of 6:26:50 hours, with a total cost of \$424.68. For the sequence P1$\rightarrow$P2$\rightarrow$P3$\rightarrow$P4, we attracted 565 crowd workers (+61) over a time span of 7:16:49 hours (+49:59 minutes), with a total cost of \$431.76 (+\$7.08). Note that these numbers are based on the same values for P1 and P2. Table~\ref{tab:pages} shows that in P1 and especially P2, more participants abandoned the session early on, while in P3 \& P4, most of the contributors completed all pages of the session. This is also seen in the higher share of incorrectly answered trial questions in P1 \& P2, and in Table~\ref{tab:pages} with the higher share of judgments achieved in P3 \& P4.

\subsubsection{Crowdsourced Classifications} \label{sec:classacc}
We found that the crowd's performance was significantly more accurate in the condition with two consecutive micro-tasks P3$\rightarrow$P4 than in the singular classification task P3${^\prime}$ ($\beta = 0.4235, SE = 0.12, z = 3.49, p < .001, OR = 1.53$), with the effect size suggesting that splitting P3 \& P4 into the two consecutive subtasks P3$\rightarrow$P4 is $1.5\times$ more likely to achieve better results. In both P3 and P3${^\prime}$, the crowd classified far fewer sentences as \textit{none} than in the gold standard, with the combined score for P3$\rightarrow$P4 being slightly above the gold standard. Although detrimental to precision, fewer classifications of \textit{none} is likely to support better recall because fewer quality-related sentences are discarded. The crowd workers in P3 were more likely to classify sentences as \textit{feature} and \textit{stability}, while crowd workers in P3${^\prime}$ chose \textit{performance} nearly twice as often as in P3. A notable similarity is that in P3, the participants classified 32.05\% of the sentences as \textit{quality}, which hardly differs from the share of 29.54\% in P3${^\prime}$ for the equivalent combination of \textit{compatibility}, \textit{user-friendliness} and \textit{security}.

The confusion matrix for P3 in Table~\ref{tab:confusion3} shows that the crowd achieved the best precision and recall on \textit{feature}, and a high recall on \textit{stability} with fairly good precision. The majority of the mistakes involved \textit{none} getting classified as \textit{quality} (17 instances) and vice versa (46 instances). Thus, it is likely that potentially irrelevant sentences belonging to the \textit{none} class are included in the \textit{quality} category that was used as input for P4, which might explain why the crowd workers in P4 classified a considerable share of items as \textit{other}. The crowd workers in P3 also frequently classified \textit{stability} and \textit{feature} (and to a lesser degree \textit{performance}) as \textit{quality}, which can be explained by the latter being a more generic class that was more likely to be selected in case of doubt. For \textit{none}, the precision was fairly good, but the recall was worst for all tags, while the precision was lowest for \textit{performance}.

\begin{table}[pos=h]
    \centering
    \caption{Confusion matrix comparing the results of the crowdsourced task of Phase P3 to the gold standard.}
    \label{tab:confusion3}
    \scriptsize
    \begin{tabular}{l|S[table-format=2.0]S[table-format=2.0]S[table-format=3.0]S[table-format=2.0]S[table-format=2.0]|S[table-format=1.0]S[table-format=1.0]|S[table-format=1.2]S[table-format=1.2]}
		\hline
		\rowcolor[HTML]{D0D0D0} \cellcolor[HTML]{E0E0E0} & \multicolumn{7}{c|}{\textbf{Gold Standard}} & & \\
		\rowcolor[HTML]{D0D0D0} 
		\cellcolor[HTML]{E0E0E0} \textbf{Crowd} & \textit{Quality} & \textit{\begin{tabular}[c]{@{}c@{}}Perfor- \\ mance\end{tabular}} & \textit{Stability} & \textit{Feature} & \textit{None} & \textit{\begin{tabular}[c]{@{}c@{}}Multiple \\ (Correct)\end{tabular}} & \textit{\begin{tabular}[c]{@{}c@{}}Multiple \\ (False)\end{tabular}} & \textbf{Precision} & \textbf{Recall} \\ \cline{2-10}
		\cellcolor[HTML]{E0E0E0} \textit{Quality}            & \cellcolor[HTML]{F0F0F0} 103 & 16 & 5  & 0  & 46 & 8 & 0 & 0.63 & 0.58 \\
		\cellcolor[HTML]{E0E0E0} \textit{Performance}        & 11  & \cellcolor[HTML]{F0F0F0} 16 & 1  & 3  & 5  & 1 & 1 & \cellcolor[HTML]{FFE0E0} 0.46 & 0.56 \\
		\cellcolor[HTML]{E0E0E0} \textit{Stability}          & 24  & 2  & \cellcolor[HTML]{F0F0F0} 65 & 2  & 12 & 6 & 2 & 0.63 & 0.88 \\
		\cellcolor[HTML]{E0E0E0} \textit{Feature}            & 21  & 0  & 1  & \cellcolor[HTML]{F0F0F0} 49 & 5  & 6 & 1 & \cellcolor[HTML]{E0FFE0} 0.67 & \cellcolor[HTML]{E0FFE0} 0.89 \\
		\cellcolor[HTML]{E0E0E0} \textit{None}               & 17  & 1  & 1  & 1  & \cellcolor[HTML]{F0F0F0} 38 & 0 & 0 & 0.66 & \cellcolor[HTML]{FFE0E0} 0.42 \\ \hline
		\cellcolor[HTML]{E0E0E0} \textit{Multiple (Correct)} & 21  & 8  & 8  & 5  & 21 & 3 & ~ & ~ & ~  \\
		\cellcolor[HTML]{E0E0E0} \textit{Multiple (False)}   & 18  & 1  & 2  & 1  & 13 & ~ & 0 & ~ & ~ \\ \hline
		\cellcolor[HTML]{D0D0D0} \textbf{Macro Average}      & \multicolumn{7}{c|}{} & 0.61 & 0.67 \\ 
		\cellcolor[HTML]{D0D0D0} \textbf{Accuracy}               & \multicolumn{7}{c|}{} & \multicolumn{2}{c}{0.63} \\ \hline
    \end{tabular}
\end{table}

Table~\ref{tab:confusion4} shows that in P4, the crowd workers performed best on \textit{compatibility}. The performance was mainly affected by the crowd workers confusing \textit{user-friendliness} with \textit{none} (22 instances) and vice versa (20 instances), in part due to misattributing sentences about connection quality. Sentences that the crowd workers in P3 failed to classify as \textit{none} continued to persist in P4 and were then often classified as \textit{user-friendliness}. Recall on \textit{security} was drastically reduced because two out of five classifications were misattributed as \textit{user-friendliness}. The poor performance on \textit{security} does not paint a reliable picture because of the low number of associated sentences in the gold standard for P3 \& P4.

\begin{table}[pos=h]
    \centering
    \caption{Confusion matrix comparing the results of the crowdsourced task of Phase P4 to the gold standard.}
    \label{tab:confusion4}
    \scriptsize
    \begin{tabular}{l|S[table-format=2.0]S[table-format=2.0]S[table-format=1.0]S[table-format=2.0]|S[table-format=1.0]S[table-format=1.0]|S[table-format=1.2]S[table-format=1.2]}
		\hline
		\rowcolor[HTML]{D0D0D0} \cellcolor[HTML]{E0E0E0} & \multicolumn{6}{c|}{\textbf{Gold Standard}} & & \\
		\rowcolor[HTML]{D0D0D0} 
		\cellcolor[HTML]{E0E0E0} \textbf{Crowd} & \textit{\begin{tabular}[c]{@{}c@{}}Compati- \\ bility\end{tabular}} & \textit{\begin{tabular}[c]{@{}c@{}}User- \\ Friendliness\end{tabular}} & \textit{Security} & \textit{None} & \textit{\begin{tabular}[c]{@{}c@{}}Multiple \\ (Correct)\end{tabular}} & \textit{\begin{tabular}[c]{@{}c@{}}Multiple \\ (False)\end{tabular}} & \textbf{Precision} & \textbf{Recall} \\ \cline{2-9}
		\cellcolor[HTML]{E0E0E0} \textit{Compatibility}      & \cellcolor[HTML]{F0F0F0} 22 & 2  & 0 & 2  & 4 & 0 & 0.87 & \cellcolor[HTML]{E0FFE0} 0.77 \\
		\cellcolor[HTML]{E0E0E0} \textit{User-Friendliness}  & 5  & \cellcolor[HTML]{F0F0F0} 40 & 2 & 22 & 3 & 0 & \cellcolor[HTML]{FFE0E0} 0.60 & 0.70 \\ 
		\cellcolor[HTML]{E0E0E0} \textit{Security}           & 0  & 0  & \cellcolor[HTML]{F0F0F0} 2 & 0  & 0 & 0 & \cellcolor[HTML]{E0FFE0} 1.00 & \cellcolor[HTML]{FFE0E0} 0.40 \\
		\cellcolor[HTML]{E0E0E0} \textit{None}               & 2  & 20 & 0 & \cellcolor[HTML]{F0F0F0} 45 & 3 & 1 & 0.68 & 0.70 \\ \hline 
		\cellcolor[HTML]{E0E0E0} \textit{Multiple (Correct)} & 1  & 11 & 0 & 11 & 1 & ~ & ~ & ~  \\
		\cellcolor[HTML]{E0E0E0} \textit{Multiple (False)}   & 0  & 0  & 1 & 0  & ~ & 0 & ~ & ~  \\ \hline
		\cellcolor[HTML]{D0D0D0} \textbf{Macro Average}      & \multicolumn{6}{c|}{} & 0.79 & 0.64 \\
		\cellcolor[HTML]{D0D0D0} \textbf{Accuracy}               & \multicolumn{6}{c|}{} & \multicolumn{2}{c}{0.72} \\ \hline
    \end{tabular}
\end{table}

\begin{figure}[pos=h]
	\centering
	\includegraphics[width=0.70\textwidth]{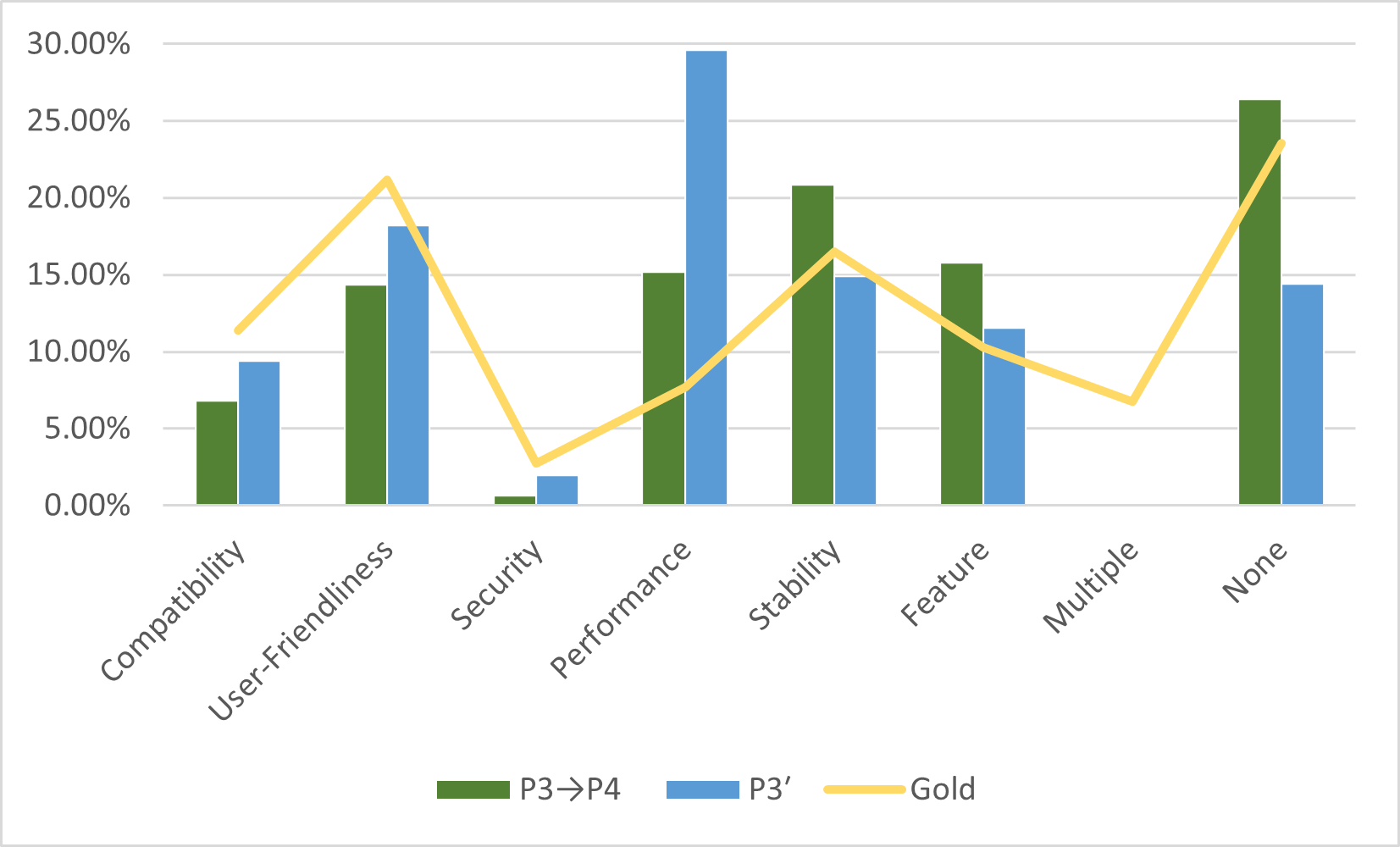}
	\caption{Distribution of the judgments assigned in P3$\rightarrow$P4 (combined score) and in P3${^\prime}$ compared to the distribution of classes in the gold standard. Individual crowd workers could not assign multiple tags. See the online appendix~\citep{appendix} for details.}
	\label{fig:distrp34}
\end{figure}

The confusion matrix for P3${^\prime}$ in Table~\ref{tab:confusion3primed} shows that the crowd in this phase performed best on \textit{stability} and \textit{feature}. The most obvious deviation from the gold standard is the large number of sentences misattributed to \textit{performance}, most of which should have been classified as \textit{user-friendliness} (48 instances) or \textit{none} (41 instances) according to the gold standard, while only 33 sentences were correctly classified as \textit{performance}. Figure~\ref{fig:distrp34} visualizes the distribution of the judgments assigned by the crowd workers in P3 \& P4, which highlights that \textit{performance} got overclassified. Several sentences that should have been classified as \textit{none} were classified as \textit{user-friendliness} (20 instances). The crowd tended to assign \textit{security} less often compared to the gold standard, possibly because its low salience in the gold standard caused its prominence to be low among crowd workers.

\begin{table}[pos=h]
    \centering
    \caption{Confusion matrix comparing the results of the crowdsourced task of Phase P3${^\prime}$ to the gold standard.}
    \label{tab:confusion3primed}
    \scriptsize
    \begin{tabular}{l|S[table-format=2.0]S[table-format=2.0]S[table-format=2.0]S[table-format=1.0]S[table-format=2.0]S[table-format=2.0]S[table-format=1.0]|S[table-format=1.0]S[table-format=1.0]|S[table-format=1.2]S[table-format=1.2]}
		\hline
		\rowcolor[HTML]{D0D0D0} \cellcolor[HTML]{E0E0E0} & \multicolumn{9}{c|}{\textbf{Gold Standard}} & & \\
		\rowcolor[HTML]{D0D0D0} 
		\cellcolor[HTML]{E0E0E0} \textbf{Crowd} & \textit{\begin{tabular}[c]{@{}c@{}}Compati- \\ bility\end{tabular}} & \textit{\begin{tabular}[c]{@{}c@{}}User- \\ Friendl.\end{tabular}} & \textit{\begin{tabular}[c]{@{}c@{}}Secu- \\ rity\end{tabular}} & \textit{\begin{tabular}[c]{@{}c@{}}Perfor- \\ mance\end{tabular}} & \textit{\begin{tabular}[c]{@{}c@{}}Stabi- \\ lity\end{tabular}} & \textit{\begin{tabular}[c]{@{}c@{}}Fea- \\ ture\end{tabular}} & \textit{None} & \textit{\begin{tabular}[c]{@{}c@{}}Mult. \\ (Corr.)\end{tabular}} & \textit{\begin{tabular}[c]{@{}c@{}}Mult. \\ (False)\end{tabular}} & \textbf{\begin{tabular}[c]{@{}c@{}}Pre- \\ cision\end{tabular}} & \textbf{\begin{tabular}[c]{@{}c@{}}Re- \\ call\end{tabular}} \\ \cline{2-12}
		\cellcolor[HTML]{E0E0E0} \textit{Compatibility} & \cellcolor[HTML]{F0F0F0} 19 & 0  & 2 & 2  & 2  & 1  & 7  & 2 & 0  & 0.58 & 0.44 \\
		\cellcolor[HTML]{E0E0E0} \textit{User-Friendl.} & 7  & \cellcolor[HTML]{F0F0F0} 34 & 2 & 1  & 1  & 3  & 20 & 1 & 1  & 0.52 & 0.43 \\
		\cellcolor[HTML]{E0E0E0} \textit{Security}      & 0  & 1  & \cellcolor[HTML]{F0F0F0} 3 & 0  & 0  & 0  & 1  & 0 & 0  & 0.71 & \cellcolor[HTML]{FFE0E0} 0.39 \\
		\cellcolor[HTML]{E0E0E0} \textit{Performance}   & 15 & 48 & 3 & \cellcolor[HTML]{F0F0F0} 33 & 12 & 6  & 41 & 8 & 10 & \cellcolor[HTML]{FFE0E0} 0.23 & \cellcolor[HTML]{E0FFE0} 0.82 \\
		\cellcolor[HTML]{E0E0E0} \textit{Stability}     & 5  & 6  & 1 & 4  & \cellcolor[HTML]{F0F0F0} 57 & 0  & 3  & 2 & 1  & \cellcolor[HTML]{E0FFE0} 0.75 & 0.76 \\
		\cellcolor[HTML]{E0E0E0} \textit{Feature}       & 4  & 3  & 0 & 1  & 1  & \cellcolor[HTML]{F0F0F0} 37 & 3  & 1 & 1  & \cellcolor[HTML]{E0FFE0} 0.75 & 0.79 \\ 
		\cellcolor[HTML]{E0E0E0} \textit{None}          & 2  & 8  & 1 & 0  & 2  & 1  & \cellcolor[HTML]{F0F0F0} 39 & 0 & 2  & 0.70 & 0.42 \\ \hline
		\cellcolor[HTML]{E0E0E0} \textit{Mult. (Corr.)} & 11 & 18 & 2 & 3  & 6  & 11 & 20 & 5 & ~  & ~ & ~ \\
		\cellcolor[HTML]{E0E0E0} \textit{Mult. (False)} & 4  & 7  & 2 & 0  & 2  & 2  & 6  & ~ & 1  & ~ & ~ \\ \hline
		\cellcolor[HTML]{D0D0D0} \textbf{Macro Avg.} & \multicolumn{9}{c|}{} & 0.61 & 0.58 \\
		\cellcolor[HTML]{D0D0D0} \textbf{Accuracy}  & \multicolumn{9}{c|}{} & \multicolumn{2}{c}{0.55} \\ \hline
    \end{tabular}
\end{table}

\subsubsection{Crowd Worker Analysis} \label{sec:agreeacc}
Table~\ref{tab:corr12} shows how often the crowd workers provided a correct response in P1 \& P2 according to the gold standards, with the largest category in each phase being where all three crowd workers gave the correct response. Although we found that P2 was slightly more difficult than P1, the agreement accuracy did not appear to have suffered.

\begin{table}[pos=h]
    \centering
    \caption{Accuracy of the crowd workers by agreement per class for Phases P1 \& P2.}
    \label{tab:corr12}
    \begin{tabular}{l|S[table-format=3.0]S[table-format=2.2]|S[table-format=3.0]S[table-format=2.2]}
		\hline
		\rowcolor[HTML]{D0D0D0} 
		\textbf{Correctness} & \multicolumn{2}{c|}{\textbf{P1 (N,\,\%)}} & \multicolumn{2}{c|}{\textbf{P2 (N,\,\%)}} \\ \hline
		3 of 3 & 684 & 70.01 & 327 & 53.87 \\ 
		2 of 3 & 175 & 17.91 & 171 & 28.17 \\ 
		1 of 3 & 77  & 7.88  & 74  & 12.19 \\ 
		0 of 3 & 41  & 4.20  & 35  & 5.77  \\ \hline
		\textbf{Total} & \textbf{977} & ~ & \textbf{607} & ~ \\
		\textbf{Average (\%)} & \textbf{84.70} & ~ & \textbf{76.72} & ~ \\ \hline
    \end{tabular}
\end{table}

Table~\ref{tab:agree34} shows the accuracy scores achieved in P3 \& P4 based on the number of crowd workers agreeing on the same class. When all six contributors agreed, the classification matched very well with the gold standard, with the lowest accuracy achieved in P3 (84.72\%). Typically, accuracy quickly decreased as agreement between crowd members decreased. The exception to this is that the accuracy for tags that two out of six crowd workers agreed on was higher than when three out of six agreed. This can be explained by a higher likelihood that two or three classes were chosen, of which at least one was the correct answer. The only occurrence of complete disagreement was observed in P3${^\prime}$, where the crowd workers assigned every possible tag except for \textit{security}. The sentence was ``If not being able to search within your notes (including text in images like you can on the pc version) is disappointing, than \textit{[sic]} the complete lack of any search feature is a disaster.'' The situation in which three out of six crowd workers agreed on one class occurred most often, which sometimes involved a tie between two classes. These were especially observed for the classification of \textit{none} and \textit{user-friendliness}, as well as \textit{none} and \textit{compatibility}. 

Table~\ref{tab:agree34} also shows that the overall accuracy of the crowd workers in P3 (63\%) outweighed that of P3${^\prime}$ (55\%), with the best results in the smaller sequential task P4 (72\%), suggesting that a multi-stage classification will lead to more accurate results. With higher agreement strongly increasing precision, the quality of the results is likely to scale with the number of classifications per sentence. Thus, a possible solution to achieving better results is not to limit the number of crowd workers per job, but to instead keep the judgment phase active until a certain number of contributors agree. 

\begin{table}[pos=h]
    \centering
    \caption{Accuracy of the crowd workers by agreement per class for Phases P3 \& P4.}
    \label{tab:agree34}
    \begin{tabular}{l|l|S[table-format=3.0]S[table-format=2.2]|S[table-format=3.0]|S[table-format=3.0]|S[table-format=1.2]}
		\hline
		\rowcolor[HTML]{D0D0D0} 
		\textbf{Phase} & \textbf{Agreement} & \multicolumn{2}{c|}{\textbf{Frequency (N,\,\%)}} & \textbf{Correct} & \textbf{Incorrect} & \textbf{Accuracy} \\ \hline
		P3            & Six out of six   & 72  & 12.61 & 61  & 11  & 0.85 \\ 
					  & Five out of six  & 117 & 20.49 & 89  & 28  & 0.76 \\ 
					  & Four out of six  & 145 & 25.39 & 91  & 54  & 0.63 \\ 
					  & Three out of six & 157 & 27.50 & 70  & 87  & 0.45 \\ 
					  & Two out of six   & 80  & 14.01 & 47  & 33  & 0.59 \\ \cline{2-7}
					  & \textbf{Total} & \textbf{571} & \textbf{100} & \textbf{358} & \textbf{213} & \textbf{0.63} \\ \hline 
		P4            & Six out of six   & 28  & 14    & 26  & 2   & 0.93 \\ 
					  & Five out of six  & 43  & 21.5  & 34  & 9   & 0.79 \\ 
					  & Four out of six  & 57  & 28.5  & 35  & 22  & 0.61 \\ 
					  & Three out of six & 68  & 34    & 44  & 24  & 0.65 \\ 
					  & Two out of six   & 4   & 2     & 4   & 0   & 1.00 \\ \cline{2-7}
					  & \textbf{Total} & \textbf{200} & \textbf{100} & \textbf{143} & \textbf{57} & \textbf{0.72} \\ \hline 
		P3${^\prime}$ & Six out of six   & 53  & 9.28  & 48  & 5   & 0.91 \\ 
					  & Five out of six  & 64  & 11.21 & 42  & 22  & 0.66 \\ 
					  & Four out of six  & 147 & 25.74 & 81  & 66  & 0.55 \\ 
					  & Three out of six & 205 & 35.9  & 77  & 128 & 0.38 \\ 
					  & Two out of six   & 101 & 17.69 & 64  & 37  & 0.63 \\ 
					  & No agreement     & 1   & 0.18  & 0   & 1   & 0.00 \\ \cline{2-7}
					  & \textbf{Total} & \textbf{571} & \textbf{100} & \textbf{312} & \textbf{259} & \textbf{0.55} \\ \hline
    \end{tabular}
\end{table}

\subsection{Results for RQ3 -- LLMs} \label{sec:results-llms}
In this section, we first provide general statistics of the experiment with LLMs (Section~\ref{sec:results-llms-stat}), before reporting on the classification performance on P1 and P2 (Section~\ref{sec:results-llms-class12}), and on P3${^\prime}$ (Section~\ref{sec:results-llms-class3}).

\subsubsection{General Statistics} \label{sec:results-llms-stat}
Overall, we used fairly long prompts for this study (M = 12,393 characters, SD = 3,556), which included the data to classify. The processing time per batch averaged 1:08 minutes (SD = 36 seconds). ChatGPT required an average of 1.76 seconds per decision, and 0.71 decisions per second. We provide detailed analyses and all data in our online appendix~\citep{appendix}. 

\subsubsection{LLM Classifications for P1 \& P2} \label{sec:results-llms-class12}

\begin{figure}[pos=h]
	\centering
	\begin{subfigure}{.5\textwidth}
		\centering
		\includegraphics[width=.95\linewidth]{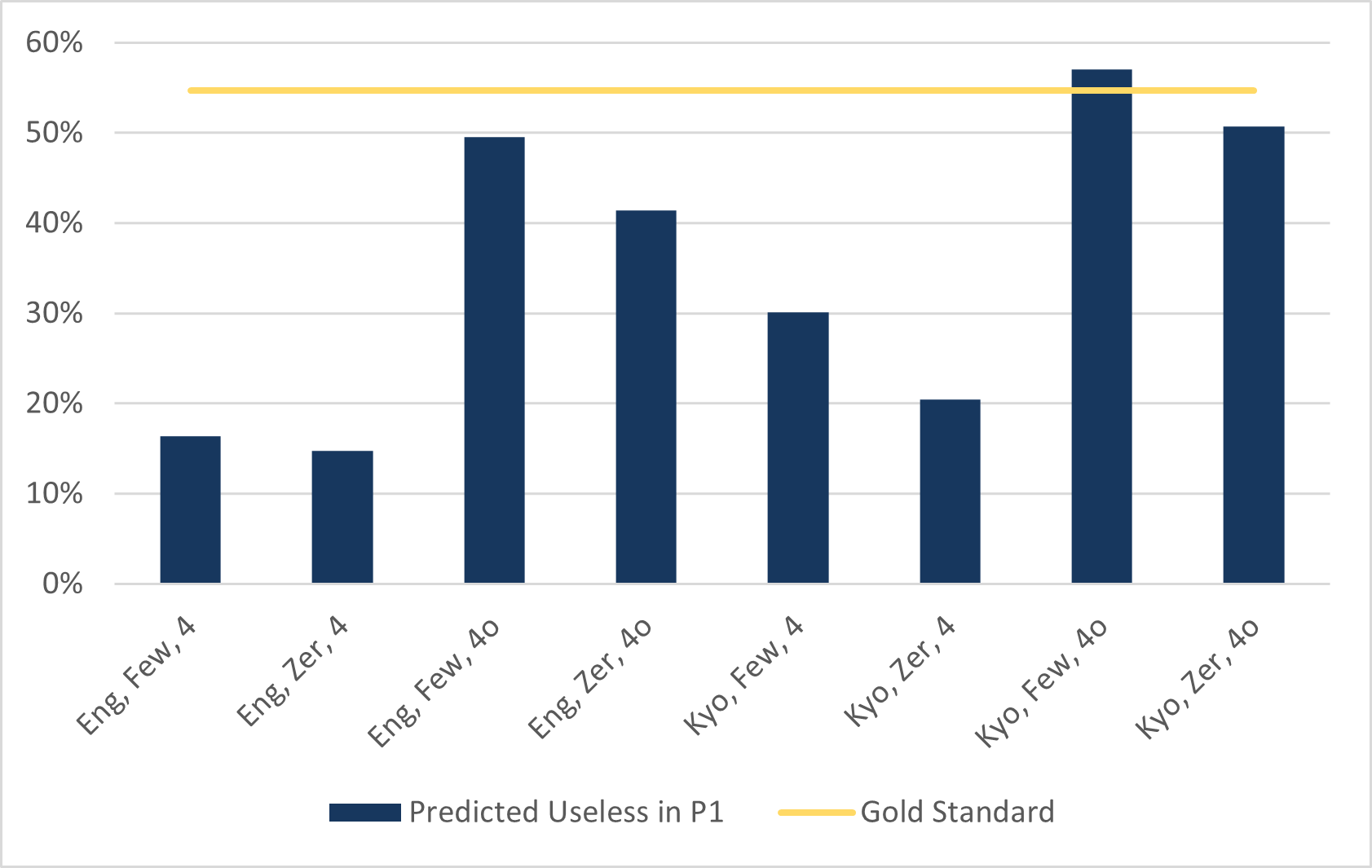}
		\caption{Distribution of tags in P1.}
		\label{fig:llms_p1_distr}
	\end{subfigure}%
	\begin{subfigure}{.5\textwidth}
		\centering
		\includegraphics[width=.95\linewidth]{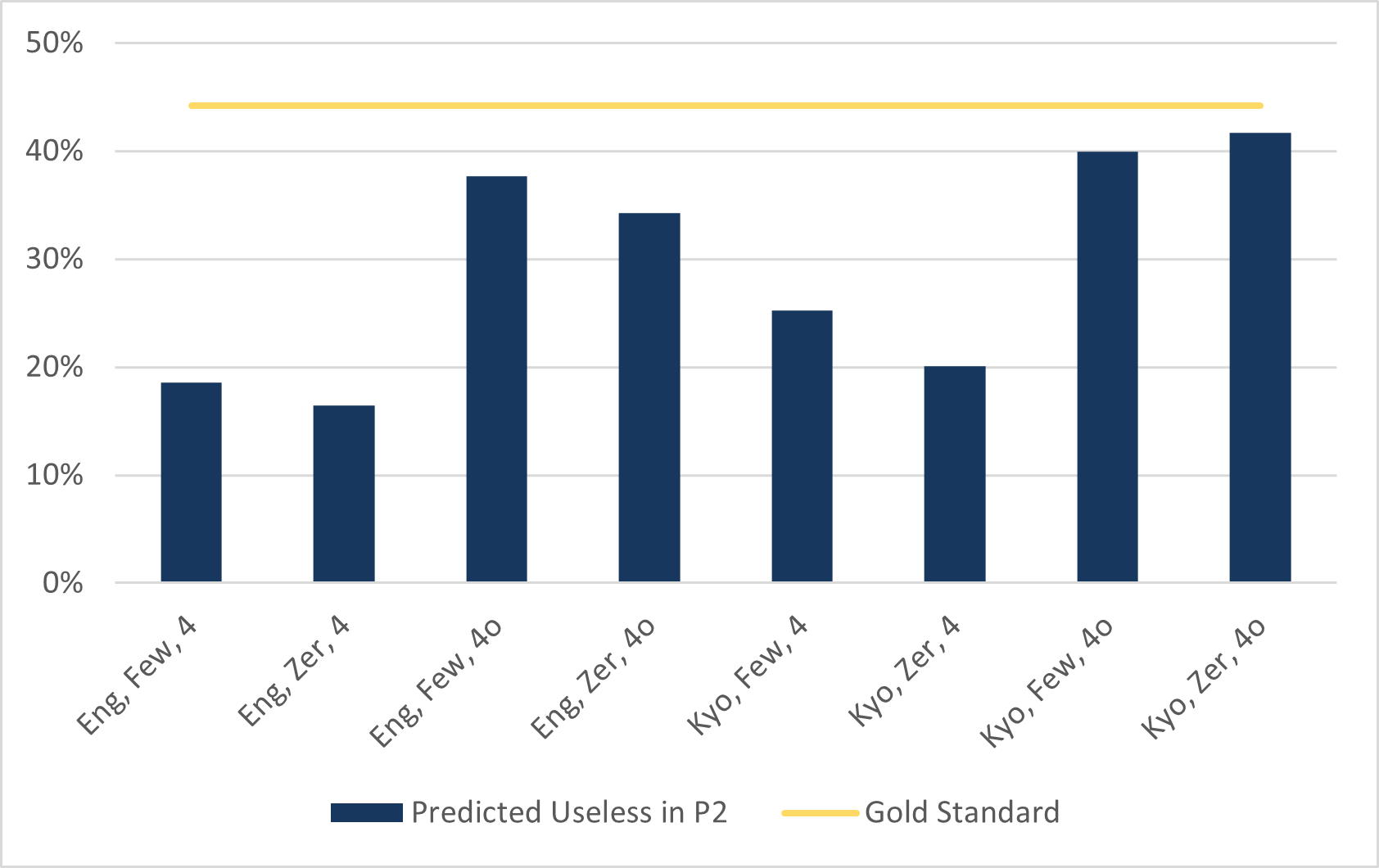}
		\caption{Distribution of tags in P2.}
		\label{fig:llms_p2_distr}
	\end{subfigure}
	\caption{Distribution of \textit{helpful} vs. \textit{useless} tags in Phases P1 and P2. The yellow line indicates the share of \textit{useless} tags in the gold standard of each phase.}
	\label{fig:llms_p12_distr}
\end{figure}

Compared to the gold standards for P1 \& P2, there is an apparent class imbalance towards classifying items as \textit{helpful}, with especially ChatGPT 4 Legacy predicting \textit{useless} far less often, as shown in Figure~\ref{fig:llms_p12_distr}. While the LLM-based classifiers achieved high recall but limited precision on \textit{helpful} in both phases, the precision for \textit{useless} was high but recall was limited. We found strong main effects and several significant interaction effects in P1 for our study’s three experimental factors:

\begin{enumerate}
    \item \textit{LLM} (4 vs. 4o): we determined a significantly better performance for ChatGPT 4o over ChatGPT 4 Legacy ($\beta = 1.04, SE = 0.10, z = 10.07, p < .001, OR = 2.83$), which suggests a clear improvement of the updated model over its predecessor, being $2.8\times$ more likely to achieve better results on this task. 
    \item \textit{Prompt type} (Eng vs. Ky\={o}: the Ky\={o}ryoku prompt outperformed the engineered prompt ($\beta = 0.52, SE = 0.10, z = 5.40, p < .001, OR = 1.67$), indicating that the structure of the prompt affects performance, with the prompt that was not specifically engineered for ChatGPT being $1.7\times$ more likely to achieve better results. 
    \item \textit{Learning strategy} (Few vs. Zer): the role of a few-shot vs. zero-shot prompting strategy had no significant effect ($\beta = -0.10, SE = 0.09, z = -1.14, p = .26, OR = 0.90$), with the odds of omitting examples to negatively affect performance being just 10\% in P1.
\end{enumerate}

\noindent In addition, we found that the impact of the learning strategy depending on the LLM used, as evidenced by a significant interaction effect for prompt type $\times$ learning strategy ($\beta = -0.29, SE = 0.13, z = -2.22, p = .03, OR = 0.75$), and a three-way effect ($\beta = 0.39, SE = 0.21, z = 1.85, p = .06, OR = 1.48$). Further analysis suggested that ChatGPT 4 Legacy possibly benefited more from the Ky\={o}ryoku prompt, while ChatGPT 4o appears to have benefited from the examples of the few-shot learning strategy for both prompt types.

\begin{figure}[pos=h]
    \centering
    \begin{subfigure}{.5\textwidth}
		\centering
		\includegraphics[width=.95\linewidth]{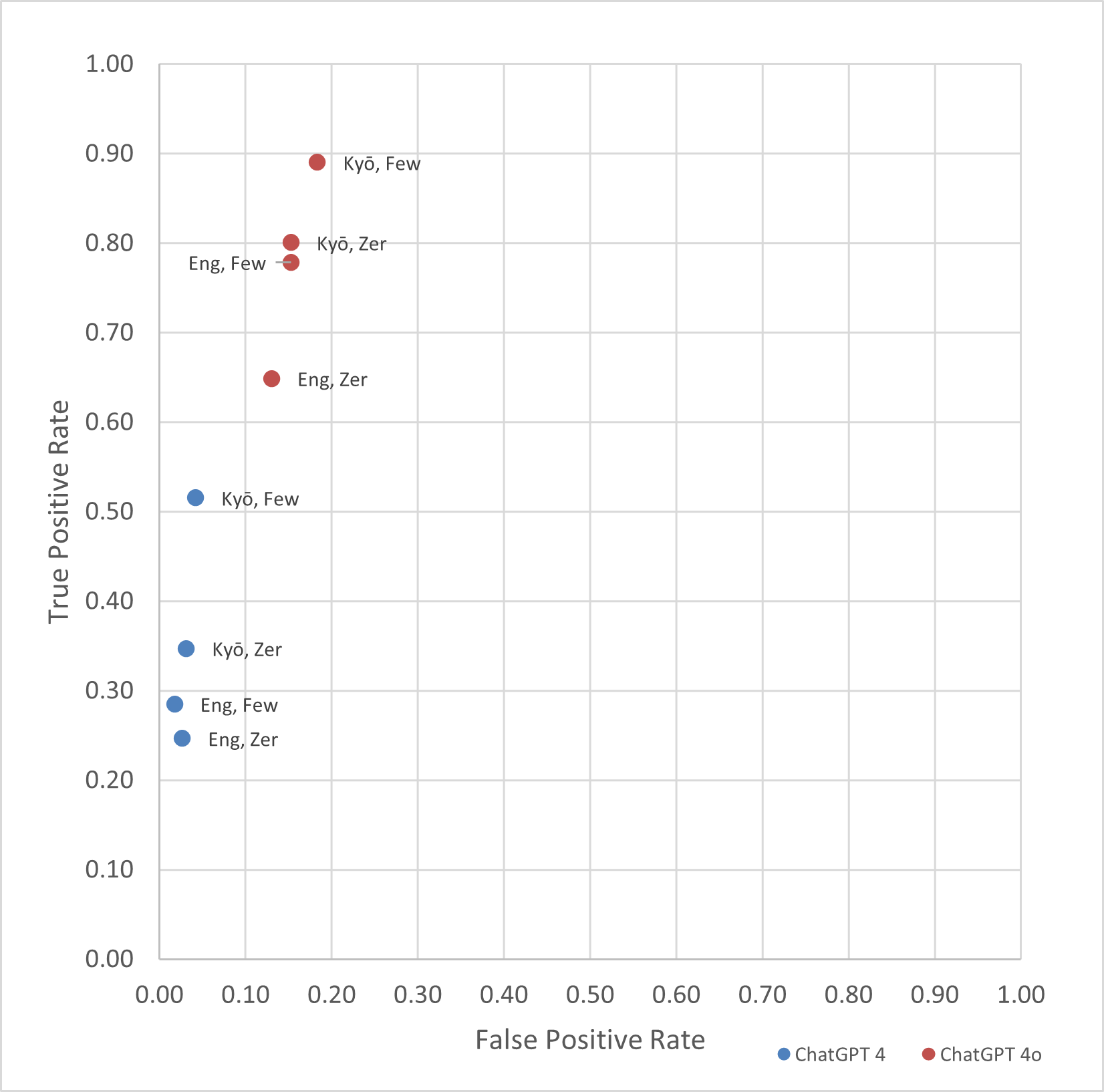}
		\caption{ROC Plot for P1.}
		\label{fig:roc_p1}
    \end{subfigure}%
    \begin{subfigure}{.5\textwidth}
		\centering
		\includegraphics[width=.95\linewidth]{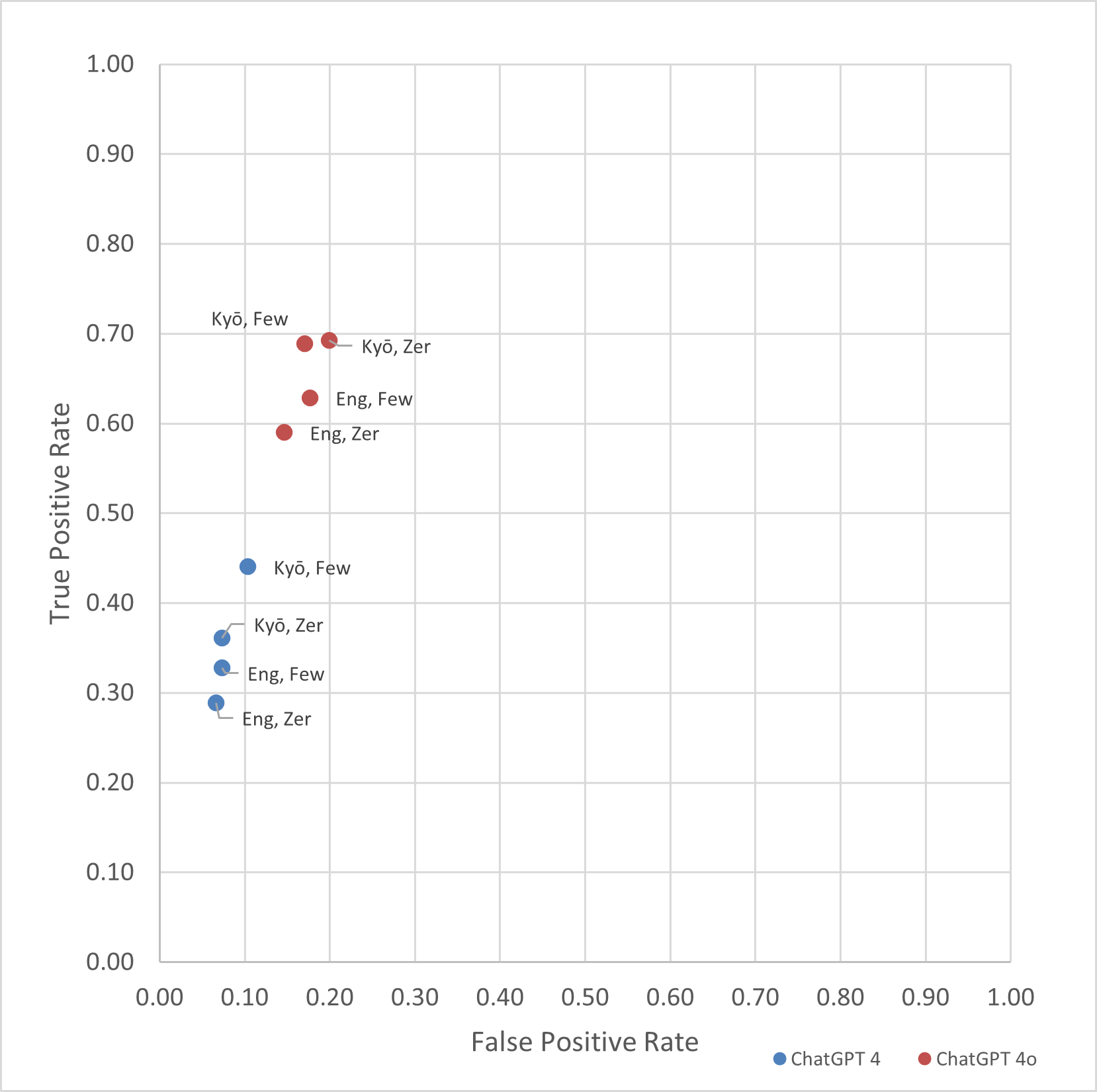}
		\caption{ROC Plot for P2.}
		\label{fig:roc_p2}
    \end{subfigure}
    \caption{Receiver Operating Characteristic (ROC) Plots for Phases P1 and P2.}
    \label{fig:test}
\end{figure}

Table~\ref{tab:llms_confusion1} shows the confusion matrix for the eight conditions in P1. ChatGPT 4 Legacy achieved better precision, while ChatGPT 4o achieved better recall. The $\lcurvyangle$Ky\={o},\,Few,\,4o$\rcurvyangle$ condition scored best with the highest recall score. The $\lcurvyangle$Eng,\,Zer,\,4$\rcurvyangle$ condition scored poorest due to limited recall. Figure~\ref{fig:roc_p1} shows that both LLMs returned few false positives (FPs), with an area under ROC curve (ROC-AUC) of 0.93 suggesting very good classifier performance in P1. However, the experimental factors LLM and prompt type appear to have an effect on classifier performance. ChatGPT 4o consistently sacrifices specificity (1 – FP rate) for sensitivity (TP rate), making it the best choice if a high TP rate is preferred (e.g., above 60\%), while ChatGPT 4 Legacy achieves higher specificity at lower sensitivity, and would be the better choice for instances where a FP rate should be low (e.g., below 10\%). Regarding prompt type, within the performance of each LLM the engineered prompts always achieved higher specificity but equal or lower sensitivity compared to the Ky\={o}ryoku prompt.

\begin{table}[pos=h]
    \centering
    \caption{Confusion matrix comparing the results of the LLMs to the gold standard for P1. The number of positives and negatives is based on the gold standard, and it reports on the ability of LLMs to correctly identify items as \textit{useless}.}
    \label{tab:llms_confusion1}
    \scriptsize
    \begin{tabular}{l|S[table-format=4.0]|S[table-format=4.0]|S[table-format=4.0]|S[table-format=4.0]|S[table-format=3.0]|S[table-format=4.0]|S[table-format=1.2]|S[table-format=1.2]}
		\hline
		\rowcolor[HTML]{D0D0D0} 
		\textbf{\begin{tabular}[c]{@{}c@{}}~ \\ Condition\end{tabular}} & \textbf{\begin{tabular}[c]{@{}c@{}}Positives: \\ Useless \\ (Gold Std.)\end{tabular}} & \textbf{\begin{tabular}[c]{@{}c@{}}Negatives: \\ Helpful \\ (Gold Std.)\end{tabular}} & \textbf{\begin{tabular}[c]{@{}c@{}}True \\ Positives\end{tabular}} & \textbf{\begin{tabular}[c]{@{}c@{}}True \\ Negatives\end{tabular}} & \textbf{\begin{tabular}[c]{@{}c@{}}False \\ Positives\end{tabular}} & \textbf{\begin{tabular}[c]{@{}c@{}}False \\ Negatives\end{tabular}} & \textbf{Precision} & \textbf{Recall}  \\ \hline
		Eng,\,Few,\,4      & 516 & 484 & 152 & 472 & 12  & 364 & \cellcolor[HTML]{E0FFE0} 0.93 & 0.29 \\
		Eng,\,Zer,\,4      & 516 & 484 & 135 & 472 & 12  & 381 & 0.92 & \cellcolor[HTML]{FFE0E0} 0.26 \\
		Eng,\,Few,\,4o     & 516 & 484 & 415 & 404 & 80  & 101 & 0.84 & 0.80 \\
		Eng,\,Zer,\,4o     & 516 & 484 & 347 & 417 & 67  & 169 & 0.84 & 0.67 \\
		Ky\={o},\,Few,\,4  & 516 & 484 & 277 & 460 & 24  & 239 & 0.92 & 0.54 \\
		Ky\={o},\,Zer,\,4  & 516 & 484 & 185 & 465 & 19  & 331 & 0.91 & 0.36 \\
		Ky\={o},\,Few,\,4o & 516 & 484 & 455 & 369 & 115 & 61  & \cellcolor[HTML]{FFE0E0} 0.80 & \cellcolor[HTML]{E0FFE0} 0.88 \\
		Ky\={o},\,Zer,\,4o & 516 & 484 & 415 & 392 & 92  & 101 & 0.82 & 0.80 \\ \hline
		\textbf{Total} & \textbf{4,128} & \textbf{3,872} & \textbf{2,381} & \textbf{3,451} & \textbf{421} & \textbf{1,747} & \textbf{0.85} & \textbf{0.58} \\ \hline
    \end{tabular}
\end{table}

In P2, we observed a much less clear pattern, with the bias towards tagging items as \textit{helpful}, being most pronounced in the $\lcurvyangle$Eng,\,Few$\rcurvyangle$ and $\lcurvyangle$Eng,\,Zer$\rcurvyangle$ conditions. ChatGPT 4o continued to score better than ChatGPT 4 Legacy ($\beta = 0.35, SE = 0.08, z = 4.24, p < .001, OR = 1.42$), but there was only a trend toward significance for the performance using the Ky\={o}ryoku prompt versus the engineered prompt ($\beta = 0.15, SE = 0.08, z = 1.82, p = .07, OR = 1.16$). Neither the learning strategy nor any of the interactions were significant.

\begin{table}[pos=h]
    \centering
    \caption{Confusion matrix comparing the the results of the LLMs to the gold standard for P2. The number of positives and negatives is based on the gold standard, and it reports on the ability of LLMs to correctly identify items as \textit{useless}.}
    \label{tab:llms_confusion2}
    \scriptsize
    \begin{tabular}{l|S[table-format=4.0]|S[table-format=4.0]|S[table-format=4.0]|S[table-format=4.0]|S[table-format=3.0]|S[table-format=4.0]|S[table-format=1.2]|S[table-format=1.2]}
		\hline
		\rowcolor[HTML]{D0D0D0} 
		\textbf{\begin{tabular}[c]{@{}c@{}}~ \\ Condition\end{tabular}} & \textbf{\begin{tabular}[c]{@{}c@{}}Positives: \\ Useless \\ (Gold Std.)\end{tabular}} & \textbf{\begin{tabular}[c]{@{}c@{}}Negatives: \\ Helpful \\ (Gold Std.)\end{tabular}} & \textbf{\begin{tabular}[c]{@{}c@{}}True \\ Positives\end{tabular}} & \textbf{\begin{tabular}[c]{@{}c@{}}True \\ Negatives\end{tabular}} & \textbf{\begin{tabular}[c]{@{}c@{}}False \\ Positives\end{tabular}} & \textbf{\begin{tabular}[c]{@{}c@{}}False \\ Negatives\end{tabular}} & \textbf{Precision} & \textbf{Recall}  \\ \hline
		Eng,\,Few,\,4      & 595 & 752 & 195 & 697 & 55  & 400 & 0.78 & 0.33 \\
		Eng,\,Zer,\,4      & 595 & 752 & 172 & 702 & 50  & 423 & 0.77 & \cellcolor[HTML]{FFE0E0} 0.20 \\
		Eng,\,Few,\,4o     & 595 & 752 & 374 & 619 & 133 & 221 & 0.74 & 0.63 \\
		Eng,\,Zer,\,4o     & 595 & 752 & 351 & 642 & 110 & 244 & 0.76 & 0.50 \\
		Ky\={o},\,Few,\,4  & 595 & 752 & 262 & 674 & 78  & 333 & 0.77 & 0.44 \\
		Ky\={o},\,Zer,\,4  & 595 & 752 & 215 & 697 & 55  & 380 & \cellcolor[HTML]{E0FFE0} 0.80 & 0.36 \\
		Ky\={o},\,Few,\,4o & 595 & 752 & 410 & 624 & 128 & 185 & 0.76 & \cellcolor[HTML]{E0FFE0} 0.69 \\
		Ky\={o},\,Zer,\,4o & 595 & 752 & 412 & 602 & 150 & 183 & \cellcolor[HTML]{FFE0E0} 0.73 & \cellcolor[HTML]{E0FFE0} 0.69 \\ \hline
		\textbf{Total} & \textbf{4,760} & \textbf{6,016} & \textbf{2,391} & \textbf{5,257} & \textbf{759} & \textbf{2,369} & \textbf{0.76} & \textbf{0.50} \\ \hline
    \end{tabular}
\end{table}

Table~\ref{tab:llms_confusion2} shows the confusion matrix for the eight conditions in P2. The performance of both LLMs decreased, but while ChatGPT 4 Legacy achieved a slightly lower precision and a similar recall compared to P1, the performance of ChatGPT 4o saw a stark decrease in both precision and recall. The $\lcurvyangle$Ky\={o},\,Few,\,4o$\rcurvyangle$ and $\lcurvyangle$Ky\={o},\,Zer,\,4o$\rcurvyangle$ conditions scored best, both with good precision and fair recall. The poorest-performing condition was $\lcurvyangle$Eng,\,Zer,\,4$\rcurvyangle$ due to limited recall, as in P1. Because P2 contained sentences from online user reviews of which at least one sentence was found \textit{helpful} in P1, the task could be considered somewhat more difficult. The ROC Plot in Figure~\ref{fig:roc_p2} reflects the decrease in performance in P2, with a lower ROC-AUC of 0.90. The difference in performance of the two LLMs and the conditions for each LLM are less distinct than in P1. However, there are similar patterns of a relatively low FP rate overall, higher sensitivity but lower specificity for ChatGPT 4o compared to ChatGPT 4 Legacy, and higher sensitivity but mostly lower specificity for the Ky\={o}ryoku prompt compared to the engineered prompt.

In Phases P1 \& P2, we see that ChatGPT 4 Legacy consistently favored classifying items as \textit{helpful} over \textit{useless}. This might suggest that ChatGPT 4 Legacy has a tendency to prioritize precision over recall, which for the \textit{useless} classification resulted in a high precision but limited recall. This effect was even more pronounced in the conditions that used the engineered prompt. In P1, ChatGPT 4 Legacy even achieved near-perfect recall on \textit{helpful}, but had only limited precision. ChatGPT 4o, on the other hand, was able to achieve a more distributed assignment of tags especially in P1, where in the $\lcurvyangle$Ky\={o},\,Few$\rcurvyangle$ and $\lcurvyangle$Ky\={o},\,Zer$\rcurvyangle$ conditions, it even classified more items as \textit{useless} than \textit{helpful}, and it achieved a score of about 0.80 on both precision and recall in P1. However, its performance deteriorated in P2, achieving a score of roughly 0.70 on average, and displaying a pattern more similar to that of ChatGPT 4 Legacy in terms of favoring classifying items as \textit{helpful} over \textit{useless}.

\subsubsection{LLM Classifications for P3${^\prime}$} \label{sec:results-llms-class3}
Table~\ref{tab:llms_p3distr} shows the distribution of judgments for the eight LLM conditions in Phase P3${^\prime}$. Compared to the gold standard, \textit{user-friendliness} and \textit{security} received far fewer classifications, and \textit{feature request} and \textit{stability} far more classifications. The classification behavior of ChatGPT 4 Legacy and ChatGPT 4o on the other classes varied greatly. ChatGPT 4 Legacy classified far more and ChatGPT 4o somewhat fewer items as \textit{none} compared to the gold standard, and ChatGPT 4 Legacy classified \textit{compatibility} less often, while ChatGPT 4o classified \textit{performance} more often.

\begin{table}[pos=h]
    \centering
    \caption{Distribution of the judgments assigned by the eight LLM conditions in P3${^\prime}$. Note that all multiple tags are counted separately, causing the sum of judgment for each condition and the gold standard to exceed 624; the number of sentences in the gold standard.}
    \label{tab:llms_p3distr}
    \resizebox{1.00\textwidth}{!}{%
    \begin{tabular}{l|S[table-format=3.0]S[table-format=3.0]S[table-format=3.0]S[table-format=3.0]S[table-format=3.0]S[table-format=3.0]S[table-format=3.0]S[table-format=3.0]S[table-format=3.0]S[table-format=3.0]}
		\hline
		\rowcolor[HTML]{D0D0D0}\textbf{Tag} & \textbf{Gold} & \textbf{Eng,\,Few,\,4} & \textbf{Eng,\,Zer,\,4} & \textbf{Eng,\,Few,\,4o} & \textbf{Eng,\,Zer,\,4o} & \textbf{Ky\={o},\,Few,\,4} & \textbf{Ky\={o},\,Zer,\,4} & \textbf{Ky\={o},\,Few,\,4o} & \textbf{Ky\={o},\,Zer,\,4o} & \textbf{Average} \\ \hline
		Compatibility     & 82  & 55  & 42  & 97  & 83  & 33  & 32  & 55  & 111 & 64  \\
		User-Friendliness & 153 & 50  & 57  & 85  & 116 & 40  & 75  & 101 & 88  & 77  \\
		Security          & 28  & 10  & 15  & 11  & 14  & 10  & 13  & 12  & 17  & 13  \\
		Performance       & 62  & 69  & 59  & 80  & 78  & 49  & 82  & 97  & 78  & 74  \\
		Stability         & 117 & 142 & 158 & 144 & 150 & 162 & 164 & 151 & 132 & 50  \\
		Feature Request   & 79  & 99  & 91  & 125 & 126 & 105 & 96  & 120 & 120 & 110 \\
		None              & 147 & 218 & 212 & 108 & 86  & 233 & 169 & 100 & 94  & 153 \\ \hline
		\textbf{Total} & \textbf{668} & \textbf{643} & \textbf{634} & \textbf{650} & \textbf{653} & \textbf{632} & \textbf{631} & \textbf{636} & \textbf{640} & \textbf{640} \\ \hline
    \end{tabular}%
    }
\end{table}

\begin{figure}[pos=h]
    \centering
    \includegraphics[width=0.75\textwidth]{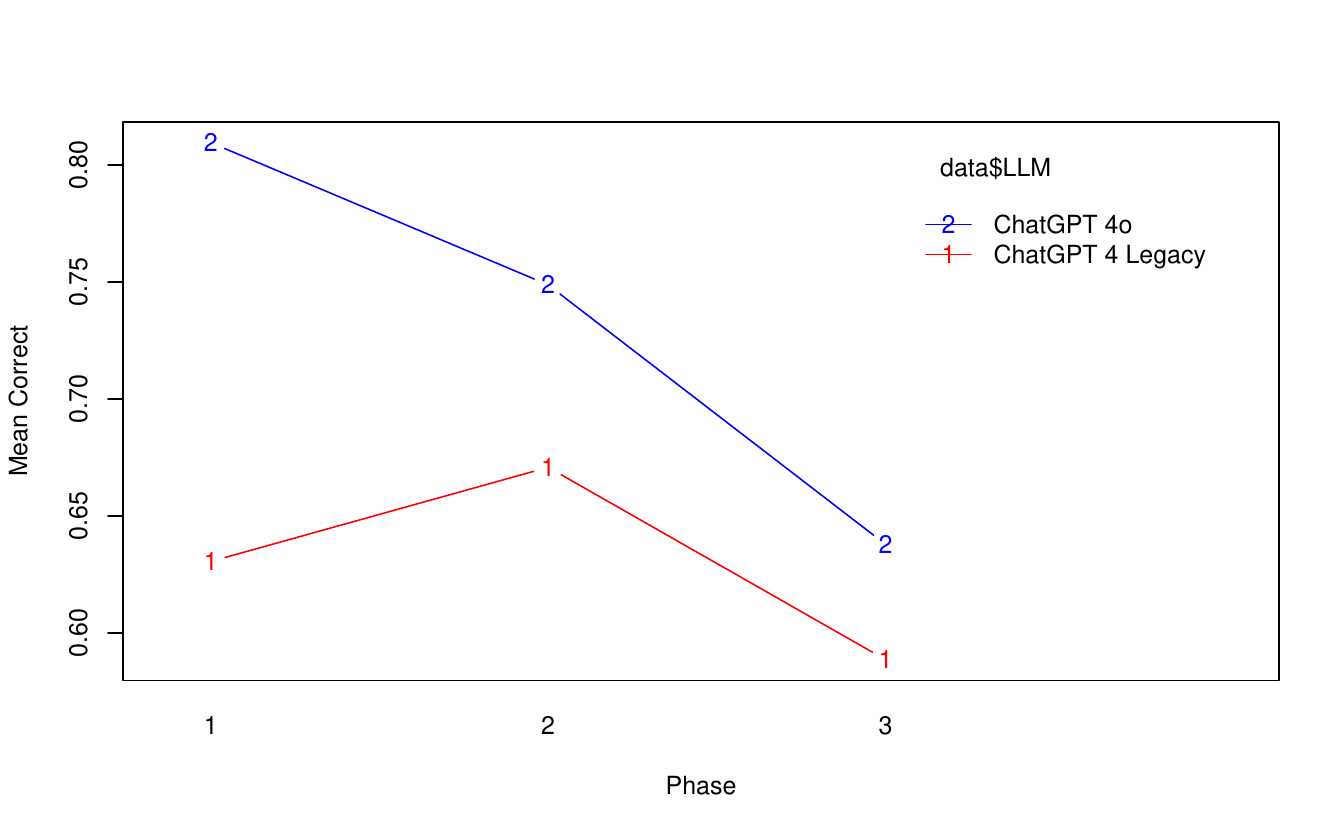}
    \caption{Interaction effect for LLM across Phases P1, P2, and P3${^\prime}$, showing a clear decrease in performance by ChatGPT 4o, and a less clear pattern but an obviously reduced performance on Phase P3${^\prime}$ by ChatGPT 4 Legacy.}
    \label{fig:llm_phases}
\end{figure}

Both ChatGPT 4 Legacy and ChatGPT 4o achieved considerably lower scores in P3${^\prime}$ compared to P1 and P2, with a significant interaction effect for phase $\times$ LLM ($\beta = -0.38, SE = 0.04, z = -9.60, p < .001, OR = 0.69$), as well as main effects for phase ($\beta = -0.06, SE = 0.03, z = -2.38, p = .02, OR = 0.94$) and LLM ($\beta = 1.23, SE = 0.08, z = 15.15, p < .001, OR = 3.43$). The effect sizes suggest that although ChatGPT 4o was $3.4\times$ more likely to perform better than ChatGPT 4 Legacy, the effect of LLM on performance decreases by 31\% in each consecutive phase, with the odds of the outcome decreasing by 6\% per phase; a number that is only this low because ChatGPT 4 Legacy performed better in P2 than in P1. Figure~\ref{fig:llm_phases} illustrates this effect, with the curve for ChatGPT 4o showing a clear decrease in performance. 

That none of the main and interaction effects in P3${^\prime}$ were significant suggests that ChatGPT 4o did not have an advantage over ChatGPT 4 Legacy in this classification task. Although the main effect for LLM did show a trend toward significance, with 19\% odds of improved performance ($\beta = 0.17, SE = 0.12, z = 1.46, p = .14, OR = 1.19$), the accuracy of the eight conditions hardly differed, ranging from 0.58 to 0.64, with both prompt type and learning strategy having negligible effects. The prediction by majority vote achieved the highest accuracy score (0.67), although its advantage over the best-performing condition was not statistically significant ($\beta = -0.14, SE = 0.12, z = -1.13, p = .26, OR = 0.87$).

Table~\ref{tab:llms_p3confusion} summarizes the performance of the eight LLM conditions. ChatGPT 4 Legacy achieved low precision on \textit{none}, among other things due to often assigning this tag to items that should be classified as \textit{user-friendliness}. ChatGPT 4o, on the other hand, often incorrectly classified sentences as \textit{compatibility}, independent of the frequency with which it assigned this tag. The prediction based on the majority vote approached or even equated the scores of the best-performing conditions while compensating for the poor predictions of the least well-performing conditions.

\begin{table}[pos=h]
    \centering
    \caption{Performance per class by the eight LLM conditions in P3${^\prime}$. The full confusion matrices were omitted for space reasons and are provided in the online appendix~\citep{appendix}. Red and green colors indicate the highest and lowest precision and recall value by class, i.e., compared horizontally.}
    \label{tab:llms_p3confusion}
    \resizebox{1.00\textwidth}{!}{%
    \begin{tabular}{l|S[table-format=1.2]S[table-format=1.2]|S[table-format=1.2]S[table-format=1.2]|S[table-format=1.2]S[table-format=1.2]|S[table-format=1.2]S[table-format=1.2]|S[table-format=1.2]S[table-format=1.2]|S[table-format=1.2]S[table-format=1.2]|S[table-format=1.2]S[table-format=1.2]|S[table-format=1.2]S[table-format=1.2]|S[table-format=1.2]S[table-format=1.2]}
		\hline
		\rowcolor[HTML]{D0D0D0} \textbf{~} & \multicolumn{2}{c|}{\textbf{Eng,\,Few,\,4}} & \multicolumn{2}{c|}{\textbf{Eng,\,Zer,\,4}} & \multicolumn{2}{c|}{\textbf{Eng,\,Few,\,4o}} & \multicolumn{2}{c|}{\textbf{Eng,\,Zer,\,4o}} & \multicolumn{2}{c|}{\textbf{Ky\={o},\,Few,\,4}} & \multicolumn{2}{c|}{\textbf{Ky\={o},\,Zer,\,4}} & \multicolumn{2}{c|}{\textbf{Ky\={o},\,Few,\,4o}} & \multicolumn{2}{c|}{\textbf{Ky\={o},\,Zer,\,4o}} & \multicolumn{2}{c}{\textbf{Majority Vote}}  \\ 
		\rowcolor[HTML]{D0D0D0} ~ & \textbf{P} & \textbf{R}  & \textbf{P} & \textbf{R} & \textbf{P} & \textbf{R} & \textbf{P} & \textbf{R} & \textbf{P} & \textbf{R} & \textbf{P} & \textbf{R} & \textbf{P} & \textbf{R} & \textbf{P} & \textbf{R} & \textbf{P} & \textbf{R} \\ \hline
		Compatibility     & 0.52 & 0.37 & 0.61 & 0.32 & 0.48 & 0.56 & 0.47 & 0.46 & 0.68 & \cellcolor[HTML]{FFE0E0} 0.27 & \cellcolor[HTML]{E0FFE0} 0.80 & 0.31 & 0.54 & 0.37 & \cellcolor[HTML]{FFE0E0} 0.46 & \cellcolor[HTML]{E0FFE0} 0.61 & 0.64 & 0.45 \\
		User-Friendliness & \cellcolor[HTML]{E0FFE0} 0.83 & 0.28 & 0.70 & 0.29 & 0.78 & 0.45 & 0.62 & \cellcolor[HTML]{E0FFE0} 0.50 & 0.80 & \cellcolor[HTML]{FFE0E0} 0.22 & \cellcolor[HTML]{FFE0E0} 0.60 & 0.31 & 0.69 & 0.46 & 0.71 & 0.43 & 0.82 & 0.45 \\
		Security          & \cellcolor[HTML]{FFE0E0} 0.60 & 0.33 & 0.64 & 0.41 & 0.64 & 0.33 & 0.67 & 0.41 & 0.70 & 0.35 & 0.69 & 0.35 & 0.67 & \cellcolor[HTML]{FFE0E0} 0.29 & 0.69 & 0.35 & \cellcolor[HTML]{E0FFE0} 
		0.82 & \cellcolor[HTML]{E0FFE0} 0.47 \\
		Performance       & 0.53 & 0.65 & 0.60 & 0.58 & 0.58 & 0.80 & 0.60 & 0.85 & \cellcolor[HTML]{E0FFE0} 0.62 & \cellcolor[HTML]{FFE0E0} 0.53 & 0.49 & 0.73 & \cellcolor[HTML]{FFE0E0} 0.48 & 0.86 & 0.60 & \cellcolor[HTML]{E0FFE0} 0.87 & 0.59 & 0.83 \\
		Stability         & 0.69 & \cellcolor[HTML]{FFE0E0} 0.84 & 0.65 & 0.89 & 0.72 & 0.88 & 0.71 & \cellcolor[HTML]{E0FFE0} 0.91 & \cellcolor[HTML]{FFE0E0} 0.62 & 0.88 & 0.62 & 0.87 & 0.68 & 0.88 & \cellcolor[HTML]{E0FFE0} 0.74 & 0.86 & 0.66 & \cellcolor[HTML]{E0FFE0} 0.91 \\
		Feature Request   & 0.66 & 0.93 & \cellcolor[HTML]{E0FFE0} 0.69 & 0.87 & \cellcolor[HTML]{FFE0E0} 0.55 & 0.91 & 0.57 & \cellcolor[HTML]{E0FFE0} 0.96 & 0.64 & 0.91 & 0.66 & \cellcolor[HTML]{FFE0E0} 0.86 & 0.56 & 0.92 & 0.57 & 0.94 & 0.68 & 0.94 \\
		None              & 0.49 & \cellcolor[HTML]{E0FFE0} 0.73 & 0.47 & 0.67 & 0.72 & 0.53 & \cellcolor[HTML]{E0FFE0} 0.78 & \cellcolor[HTML]{FFE0E0} 0.46 & \cellcolor[HTML]{FFE0E0} 0.45 & 0.71 & 0.54 & 0.62 & \cellcolor[HTML]{E0FFE0} 0.78 & 0.53 & 0.77 & 0.49 & 0.66 & 0.67 \\ \hline
		\textbf{Macro Average} & \textbf{0.62} & \textbf{0.59} & \textbf{0.62} & \textbf{0.58} & \textbf{0.64} & \textbf{0.64} & \textbf{0.63} & \textbf{0.65} & \textbf{0.64} & \textbf{0.55} & \textbf{0.63} & \textbf{0.58} & \textbf{0.63} & \textbf{0.62} & \textbf{0.65} & \textbf{0.65} & \textbf{0.69} & \textbf{0.67} \\
		\textbf{Accuracy} & \multicolumn{2}{c}{\textbf{0.60}} & \multicolumn{2}{c}{\textbf{0.59}} & \multicolumn{2}{c}{\textbf{0.64}} & \multicolumn{2}{c}{\textbf{0.64}} & \multicolumn{2}{c}{\textbf{0.58}} & \multicolumn{2}{c}{\textbf{0.59}} & \multicolumn{2}{c}{\textbf{0.63}} & \multicolumn{2}{c}{\textbf{0.64}} & \multicolumn{2}{c}{\textbf{0.67}} \\ \hline
    \end{tabular}%
    }
\end{table}

On individual classes, the classifiers were the best at predicting \textit{stability}, with fair precision and good recall, and \textit{feature request} with fair precision and good recall caused by overclassification of this class. Recall was the greatest detrimental factor on the worst-performing tags\textemdash \textit{compatibility}, \textit{user-friendliness}, and \textit{security}\textemdash  due to underclassification (with the exception of ChatGPT 4o on \textit{compatibility}). \textit{Compatibility} was frequently misclassified as \textit{none} by ChatGPT 4 Legacy and as \textit{user-friendliness} or \textit{feature request} by ChatGPT 4o; \textit{user-friendliness} was misclassified into several different tags, and \textit{security} was often misclassified as \textit{stability}. Given a similar observation regarding \textit{user-friendliness} with RQ2, this could suggest that the class may be more difficult to distinguish from other classes, the definition we provided is insufficient, or the LLMs construed a notion of \textit{user-friendliness} that deviates from the definition and examples we provided. The majority vote prediction on individual classes was equal or superior to the best-performing condition. While~\citet{Mizrahi24} recommend to average the output using multiple prompts, which inevitably causes the highest scores to be averaged out by the lowest scores, our evidence suggests that a majority vote has the potential to compensate for the poor precision or recall of individual conditions.

Table~\ref{tab:llms_p3agree} shows the accuracy of the LLMs based on the similarity between judgments. Note that here, we are comparing eight conditions with three different experimental factors; prompt type, learning strategy, and LLM. The agreement between multiple instances of the same condition would likely have been higher. The large share of perfect and near-perfect agreement suggests strong coherence between instances of ChatGPT despite different configurations. When all eight classifiers were in full agreement, they were accurate in 85.5\% of the cases. Interestingly, accuracy quickly decreased, with three to six out of eight judgments in agreement all being at or just below 50\% accuracy. The one outlier was the sentence ``If you upgrade your power-ups to level 8, you will not be able to add them to your current loadout.'' It had several different judgments given by a maximum of two classifiers, of which one answer had been correct, causing the judgment to be counted as correct.

\begin{table}[pos=h]
    \centering
    \caption{Accuracy of the LLM conditions by agreement on one class for P3${^\prime}$.}
    \label{tab:llms_p3agree}
    \begin{tabular}{l|S[table-format=3.0]S[table-format=3.2]|S[table-format=3.0]|S[table-format=3.0]|S[table-format=1.2]}
		\hline
		\rowcolor[HTML]{D0D0D0} \textbf{Agreement} & \multicolumn{2}{c|}{\textbf{ Freq (N,~\%)}} & \textbf{Correct} & \textbf{Incorrect} & \textbf{Accuracy}  \\ \hline
		Eight out of eight   & 269 & 43.11 & 230 & 39 & 0.86 \\
		Seven out of eight   & 63  & 10.10 & 46  & 17 & 0.73 \\
		Six out of eight     & 72  & 11.54 & 36  & 36 & 0.50 \\
		Five out of eight    & 109 & 17.47 & 53  & 56 & 0.49 \\
		Four out of eight    & 85  & 13.62 & 42  & 43 & 0.49 \\
		Three out of eight   & 25  & 4.01  & 12  & 13 & 0.48 \\
		Two out of eight     & 1   & 0.16  & 1   & 0  & 1.00 \\ \hline
		\textbf{Total} & \textbf{624} & \textbf{100.00} & \textbf{420} & \textbf{204} & \textbf{0.67} \\ \hline   
    \end{tabular}
\end{table}

\subsection{Results for RQ4 -- Comparison} \label{sec:results-synth}
The crowd workers and ChatGPT Best (the best-performing ChatGPT condition) in P1 $\lcurvyangle$Ky\={o},\,Few,\,4o$\rcurvyangle$ performed similarly well ($\beta = -0.21, SE = 0.13, z = -1.59, p = .25, OR = 0.81$) and significantly better than ChatGPT Majority Vote, with the crowdsourced prediction being $2.1\times$ more likely to perform better ($\beta = -0.75, SE = 0.13, z = -5.97, p < .001, OR = 0.47$), and ChatGPT Best $1.7\times$ ($\beta = -0.54, SE = 0.12, z = -4.50, p < .001, OR = 0.58$). The precision and recall scores for the three classifier conditions for P1 in Table~\ref{tab:synth_confusion_p12} show that ChatGPT Majority Vote outperformed the crowd on precision for \textit{helpful} and recall for \textit{useless}, but that its other scores did not match those of the other conditions.

\begin{table}[pos=h]
    \centering
    \caption{Performance per class for the crowdsourced tasks and LLM conditions compared to the gold standards for P1 \& P2. Green colors indicate the highest precision and recall value by class and phase, i.e., compared horizontally for P1 \& P2.}
    \label{tab:synth_confusion_p12}
    \begin{tabular}{l|S[table-format=1.2]S[table-format=1.2]|S[table-format=1.2]S[table-format=1.2]|S[table-format=1.2]S[table-format=1.2]||S[table-format=1.2]S[table-format=1.2]|S[table-format=1.2]S[table-format=1.2]|S[table-format=1.2]S[table-format=1.2]}
		\hline
		\rowcolor[HTML]{D0D0D0} \textbf{~} & \multicolumn{6}{c||}{\textbf{Phase P1}} & \multicolumn{6}{c}{\textbf{Phase P2}} \\ 
		\rowcolor[HTML]{D0D0D0} \textbf{~} & \multicolumn{2}{c|}{\textbf{Crowd}} & \multicolumn{2}{c|}{\textbf{ChatGPT Best}} & \multicolumn{2}{c||}{\textbf{ChatGPT Maj.}} & \multicolumn{2}{c|}{\textbf{Crowd}} & \multicolumn{2}{c|}{\textbf{ChatGPT Best}} & \multicolumn{2}{c}{\textbf{ChatGPT Maj.}}  \\ 
		\rowcolor[HTML]{D0D0D0} ~ & \textbf{P} & \textbf{R}  & \textbf{P} & \textbf{R} & \textbf{P} & \textbf{R} & \textbf{P} & \textbf{R} & \textbf{P} & \textbf{R} & \textbf{P} & \textbf{R}  \\ \hline
		Helpful & 0.93 & 0.83 & 0.82 & \cellcolor[HTML]{E0FFE0} 0.86 & \cellcolor[HTML]{E0FFE0} 0.96 & 0.70 & 0.91 & \cellcolor[HTML]{E0FFE0} 0.85 & 0.83 & 0.77 & \cellcolor[HTML]{E0FFE0} 0.92 & 0.71 \\
		Useless & 0.84 & 0.93 & \cellcolor[HTML]{E0FFE0} 0.89 & 0.85 & 0.60 & \cellcolor[HTML]{E0FFE0} 0.94 & \cellcolor[HTML]{E0FFE0} 0.80 & \cellcolor[HTML]{E0FFE0} 0.87 & 0.69 & 0.76 & 0.48 & 0.81 \\ \hline
		\textbf{Macro Average} & \textbf{0.89} & \textbf{0.88} & \textbf{0.85} & \textbf{0.86} & \textbf{0.78} & \textbf{0.82} & \textbf{0.85} & \textbf{0.86} & \textbf{0.76} & \textbf{0.77} & \textbf{0.70} & \textbf{0.76} \\
		\textbf{Accuracy} & \multicolumn{2}{c|}{\textbf{0.88}} & \multicolumn{2}{c|}{\textbf{0.86}} & \multicolumn{2}{c||}{\textbf{0.78}} & \multicolumn{2}{c|}{\textbf{0.86}} & \multicolumn{2}{c|}{\textbf{0.77}} & \multicolumn{2}{c}{\textbf{0.73}} \\ \hline 
    \end{tabular}%
\end{table}

In P2, the performance of the two best ChatGPT conditions was very similar, but our analysis showed that $\lcurvyangle$Ky\={o},\,Few,\,4o$\rcurvyangle$ had a slightly higher mean score than $\lcurvyangle$Ky\={o},\,Zer,\,4o$\rcurvyangle$. In this phase, the crowd achieved significantly higher scores compared to ChatGPT, being $1.8\times$ more likely to perform better than ChatGPT Best ($\beta = -0.60, SE = 0.10, z = -5.91, p < .001, OR = 0.55$), and $2.2\times$ better than ChatGPT Majority Vote ($\beta = -0.78, SE = 0.10, z = -7.75, p < .001, OR = 0.46$). The precision and recall scores for the three classifier conditions for P2 in Table~\ref{tab:synth_confusion_p12} show that the crowd outperformed ChatGPT overall. Despite its poor precision on \textit{useless}, ChatGPT Majority Vote was not significantly worse compared to ChatGPT Best, although we did observe a trend toward significance ($\beta = -0.18, SE = 0.09, z = -2.03, p = .11, OR = 0.83$), particularly due to the former's poor precision on \textit{useless}.

Table~\ref{tab:synth_confusion_p3} shows the precision and recall scores for the five classifier pipelines considered for our comparison in P3 \& P4. We excluded the \textit{feature request} class because it was not part of the LP experiment of RQ1 and does not constitute a quality characteristic. Although this did slightly change the precision and recall values compared to the data reported above, it allows for a fairer comparison. With the accuracy ranging from 0.41 to 0.68, the effect of classifier was obviously significant ($\beta = 0.22, SE = 0.03, z = 8.08, p < .001, OR = 1.24$). Unsurprisingly, Tukey’s post-hoc test showed that the classification based on the LP-based approach was significantly worse ($p<0.001$) compared to all other classifiers. With RQ2, we had already established that Crowd P3$\rightarrow$P4 had achieved statistically significantly better results than Crowd P3${^\prime}$, but Crowd P3${^\prime}$ is also outperformed by ChatGPT Best ($\beta = 0.40, SE = 0.12, z = 3.36, p = .007, OR = 1.49$) and ChatGPT Majority Vote ($\beta = 0.53, SE = 0.12, z = 4.45, p < .001, OR = 1.70$). With RQ3, we determined that ChatGPT Majority Vote was not statistically significantly different from ChatGPT Best. In this comparison, we also observed no statistically significant difference of both these results compared to the crowd in Phases P3$\rightarrow$P4. This shows that in terms of overall performance, the three best classifiers do not differ from one another.

\begin{table}[pos=h]
    \centering
    \caption{Performance per class by the five selected conditions for RQ4 compared to the gold standard for P3 \& P4. Red and green colors indicate the highest and lowest precision and recall value by class, i.e., compared horizontally.}
    \label{tab:synth_confusion_p3}
    \begin{tabular}{l|S[table-format=1.2]S[table-format=1.2]|S[table-format=1.2]S[table-format=1.2]|S[table-format=1.2]S[table-format=1.2]|S[table-format=1.2]S[table-format=1.2]|S[table-format=1.2]S[table-format=1.2]}
		\hline
		\rowcolor[HTML]{D0D0D0} ~ & \multicolumn{2}{c|}{\begin{tabular}[c]{@{}cl@{}}\textbf{Language} \\ \textbf{Patterns}\end{tabular}} & \multicolumn{2}{c|}{\begin{tabular}[c]{@{}c@{}}\textbf{Crowd} \\ \textbf{P3$\rightarrow$P4}\end{tabular}} & \multicolumn{2}{c|}{\begin{tabular}[c]{@{}c@{}}\textbf{Crowd} \\ \textbf{P3${^\prime}$} \end{tabular}} & \multicolumn{2}{c|}{\begin{tabular}[c]{@{}c@{}}\textbf{ChatGPT} \\ \textbf{Best}\end{tabular}} & \multicolumn{2}{c}{\begin{tabular}[c]{@{}c@{}}\textbf{ChatGPT} \\ \textbf{Maj. Vote}\end{tabular}}  \\
		\rowcolor[HTML]{D0D0D0} ~ & \textbf{P} & \textbf{R}  & \textbf{P} & \textbf{R} & \textbf{P} & \textbf{R} & \textbf{P} & \textbf{R} & \textbf{P} & \textbf{R}  \\ \hline
		Compatibility     & \cellcolor[HTML]{FFE0E0} 0.40 & \cellcolor[HTML]{FFE0E0} 0.06 & 0.54 & \cellcolor[HTML]{E0FFE0} 0.87 & 0.60 & 0.47 & 0.47 & 0.67 & \cellcolor[HTML]{E0FFE0} 0.68 & 0.48 \\
		User-Friendliness & 0.81 & \cellcolor[HTML]{FFE0E0} 0.10 & \cellcolor[HTML]{FFE0E0} 0.50 & \cellcolor[HTML]{E0FFE0} 0.61 & 0.54 & 0.44 & 0.71 & 0.52 & \cellcolor[HTML]{E0FFE0} 0.82 & 0.49 \\
		Security          & 0.42 & \cellcolor[HTML]{FFE0E0} 0.24 & \cellcolor[HTML]{FFE0E0} 0.20 & \cellcolor[HTML]{E0FFE0} 1.00 & 0.71 & 0.39 & 0.69 & 0.43 & \cellcolor[HTML]{E0FFE0} 0.82 & 0.53 \\
		Performance       & 0.47 & \cellcolor[HTML]{FFE0E0} 0.13 & 0.57 & 0.49 & \cellcolor[HTML]{FFE0E0} 0.24 & 0.84 & \cellcolor[HTML]{E0FFE0} 0.60 & \cellcolor[HTML]{E0FFE0} 0.87 & 0.59 & 0.83 \\
		Stability         & \cellcolor[HTML]{E0FFE0} 0.97 & \cellcolor[HTML]{FFE0E0} 0.50 & 0.90 & 0.64 & 0.75 & 0.77 & 0.74 & 0.88 & \cellcolor[HTML]{FFE0E0} 0.66 & \cellcolor[HTML]{E0FFE0} 0.92 \\
		None              & \cellcolor[HTML]{FFE0E0} 0.31 & \cellcolor[HTML]{E0FFE0} 0.95 & 0.70 & 0.62 & 0.71 & \cellcolor[HTML]{FFE0E0} 0.40 & \cellcolor[HTML]{E0FFE0} 0.77 & 0.53 & 0.66 & 0.71 \\ \hline
		\textbf{Macro Average} & \textbf{0.56} & \textbf{0.33} & \textbf{0.57} & \textbf{0.71} & \textbf{0.59} & \textbf{0.55} & \textbf{0.66} & \textbf{0.65} & \textbf{0.71} & \textbf{0.66} \\
		\textbf{Accuracy} & \multicolumn{2}{c}{\textbf{0.41}} & \multicolumn{2}{c}{\textbf{0.64}} & \multicolumn{2}{c}{\textbf{0.53}} & \multicolumn{2}{c}{\textbf{0.66}} & \multicolumn{2}{c}{\textbf{0.68}} \\ \hline 
    \end{tabular}%
\end{table}

Table~\ref{tab:synth_classes} summarizes the results of our per-class comparisons for P3 \& P4. There were significant differences between the conditions for \textit{user-friendliness}, \textit{performance}, \textit{stability}, and \textit{none}, and a modest significance for \textit{compatibility}. The findings per class are as follows:

\begin{itemize}
    \item Pairwise comparisons for \textit{user-friendliness} showed that ChatGPT Majority Vote differed significantly from all classifiers except ChatGPT Best, and that the poorest-performing classifier, Crowd P3${^\prime}$, scored significantly lower than both ChatGPT classifiers. 
    \item For \textit{performance}, the difference between classifiers was the most pronounced of all classes because Crowd P3${^\prime}$ scored considerably lower than all other classifiers; ChatGPT Best was better than both crowd classifiers. 
    \item For \textit{stability}, Crowd P3$\rightarrow$P4 had significant pairwise differences compared to all other classifiers. 
    For \textit{none}, the difference between classifiers was very pronounced because the worst-performing classifier, Language Patterns, scored significantly lower than all other classifiers. ChatGPT Best differed significantly from Language Patterns and Crowd P3$\rightarrow$P4.
    \item For \textit{compatibility}, none of the pairwise comparisons were significant, and the prediction by Crowd P3$\rightarrow$P4 did not differ significantly from the mean. 
    \item \textit{Security} was the only class for which there were no significant differences, and the prediction by ChatGPT Majority Vote prediction was also not significantly different from the other classifiers. 
\end{itemize}
\begin{table}[pos=h]
    \centering
    \caption{Statistics for the performance of the five selected conditions for RQ4} per class for P3${^\prime}$. The Friedman Test ($\chi^2; df = 4, N = 511$) indicates whether or not there was a significant difference between classifiers; the binomial GLM ($\beta$) measures whether the highest mean reported in the table was significantly different from the average for all classifiers.
    \label{tab:synth_classes}
    \begin{tabular}{l|S[table-format=3.1]S[table-format=2.3]|l|S[table-format=1.2]S[table-format=1.2]S[table-format=1.2]S[table-format=2.3]S[table-format=1.2]}
		\hline
		\rowcolor[HTML]{D0D0D0} \textbf{Class} & \textbf{$\chi^2$} & \textbf{$p$} & \textbf{Highest Mean} & \textbf{$\beta$} & \textbf{$SE$} & \textbf{$z$} & \textbf{$p$} & \textbf{$OR$} \\ \hline
		Compatibility     & 13.0  & \num{< .01}  & Crowd P3$\rightarrow$P4 & 0.26 & 0.15 & 1.80 & 0.07          & 1.30 \\
		User-Friendliness & 29.5  & \num{< .001} & ChatGPT Maj.   & 0.40 & 0.12 & 3.40 & \num{< .001} & 1.49 \\
		Security          & 4.2   & .374         & ChatGPT Maj.   & 0.23 & 0.24 & 0.95 & 0.34         & 1.26 \\
		Performance       & 265.0 & \num{< .001} & ChatGPT Best   & 0.97 & 0.16 & 5.95 & \num{< .001} & 2.63 \\
		Stability         & 28.5  & \num{< .001} & ChatGPT Best   & 0.38 & 0.15 & 2.48 & \num{< .01}  & 1.47 \\
		None              & 253.0 & \num{< .001} & ChatGPT Best   & 0.81 & 0.12 & 6.97 & \num{< .001} & 2.25 \\ \hline
    \end{tabular}
\end{table}

\vfill
\newpage

\section{Discussion of Findings} \label{sec:discussion}
This section discusses the implications of our findings according to the four research questions.

    \begin{enumerate}[\bfseries RQ1.]
        \item \textbf{How to adapt the NFR Method in order to derive LPs for the identification of quality characteristics in online user feedback?}
    \end{enumerate}
\noindent To answer RQ1, we determined the feasibility of adapting the NFR Method~\citep{Doerr11}, and in a sub-question the accuracy of the outcomes of applying LPs. Regarding the adaptation, we obtained mixed results. Expert workshops were a suitable means of eliciting a substantial number of K\&Ps related to quality characteristics and subcharacteristics\textemdash a level of granularity that research works usually do not address~\citep{Groen17users}. We also managed to encode LPs from the K\&Ps. Often, several K\&Ps were combined into one LP, but because negations and antonyms were placed into different LPs, about as many LPs were created as the number of K\&P pairs that were obtained. However, LPs have been found to be inefficient because the manual vetting process involves substantial human labor, which outweighs any benefits of reusability and potentially high precision. In addition, they achieved low recall, partially because the coverage of relevant content based on the K\&Ps elicited from just one workshop is limited. 

\textbf{RQ1a} assessed the effectiveness of the LPs based on quality-related keywords in identifying quality characteristics. The initial set of LPs achieved an average precision of $65\%$, which increased to $87\%$ after refinements were made based on the seen data. Our validation over an independent test set showed a drastic reduction in precision to 56\% on average, with two classes retaining high precision. Despite the fact that the LPs maintained good precision on \textit{usability}, they achieved the second-lowest recall score on this class. However, \textit{reliability} achieved near-perfect precision with good recall. This is an important finding, because reliability issues often impact larger groups of end-users; sentences pertaining to this class mostly report on the app crashing, freezing, hanging, breaking, or quitting. Tested against our gold standard, the LP-based approach typically achieved low recall, which suggests that LPs are very selective in their prediction of quality characteristics. This is due to the heterogeneity of online user feedback in which laypeople usually do not describe the circumstances in sufficient detail. Moreover, an approach that rigidly matches a combination of words is unable to infer cues from the context that could help resolve ambiguities, and consequently will have difficulty correctly determining the affected quality. For example, the phrase ``app doesn't work'' could imply a problem with \textit{installability}, \textit{reliability}, \textit{compatibility}, \textit{interoperability} or \textit{security}, or maybe reflect the user failing to understand how to use the app. This may also be due to end-users perceiving the software exclusively from a front-end perspective. 

\begin{mdframed}
    \textbf{Answer to RQ1:}
    Expert workshops are suitable to produce a keyword meta-model for software product quality characteristics, but the encoding of language patterns is a time-intensive task. The performance of the LPs when classifying online user feedback was poor, and was only able to classify sentences about \textit{reliability} fairly well.
\end{mdframed}

\vspace{1em}
    \begin{enumerate}[\bfseries RQ2.]
        \item \textbf{How to design micro-tasks so that lay workers through crowdsourcing can achieve good results classifying quality aspects in online user feedback?}
    \end{enumerate}
\noindent For RQ2, we extended the research on Ky\={o}ryoku~\citep{Vliet20}, a crowdsourcing method for eliciting user requirements from online user feedback. We investigated how the decomposition of a classification task into smaller, consecutive classification tasks affects the performance of crowd workers (\textbf{RQ2a}). In a single-group experiment, we attracted a crowd of of 711 non-expert human annotators, of whom 555 eventually contributed to the classification of a random sample of 1,000 app store reviews into requirements-relevant dimensions and quality aspects. We compared two conditions: one in which we presented all quality aspects at once (Phase P3${^\prime}$), and one in which we split the classification process into two micro-tasks with fewer quality aspects (Phases P3$\rightarrow$P4).

We found that the crowd workers performed reasonably well on this complex task in the decomposed P3$\rightarrow$P4 sequence, achieving an accuracy of 63\% and 72\%. The crowd workers in P3${^\prime}$ performed considerably worse, with an accuracy of 55\%. Decomposing the classification problem into several simpler tasks seems to lead to better crowd performance. However, there is also a downside to performing multiple tasks in sequence: misclassified sentences (i.e., reduced precision) on a class are prevented from being correctly classified in a later phase, at the detriment of recall for the phases combined. In this study, the sentences that the crowd did not classify as \textit{quality} in P3 (see Table~\ref{tab:confusion3}) would not proceed to P4. In P3$\rightarrow$P4, the severity of this problem appears to have been mitigated by a lower tendency to classify sentences as \textit{none}, allowing more sentences to be classified in P4.

The observation that the larger number of tagging options in P3${^\prime}$ negatively impacted the crowd’s accuracy is in line with general research on manual annotation~\citep{Bayerl11}. Particularly, the broad spectrum of choices added confusion and resulted in crowd members defaulting to one tag when in doubt\textemdash specifically for \textit{performance}\textemdash reducing precision. Further analysis suggests that the crowd workers often made interpretations beyond the given definitions in the instructions. It appears that these interpretations particularly came into play for ambiguous sentences, in case of multiple statements, contradictions, or unclear language. We also found evidence of this in the crowd workers' responses to our test questions, some of whom defended their dissenting judgment. This does not only occur among laypeople; while devising the gold standards, the researchers also observed how personal heuristics interfered with their classifications. 

Crowd workers were generally able to differentiate between software product qualities represented by their classes \textbf{(RQ2b)}, even when some occasionally got confused. The frequency and accuracy of each assigned class differed between phases, usually with different classes getting confused. Instances with perfect agreement corresponded with high accuracy scores. Although it is likely that the highest agreement is achieved on the least ambiguous sentences, a configuration requiring a minimum number of crowd workers agreeing with one another might lead to more accurate results. Like other studies~\citep[e.g.,][]{Schenk11,Gilardi23}, we found that the performance of crowd workers deteriorated as task complexity increased. It nevertheless is possible to crowdsource complex tasks that could even be challenging to expert judges, with the best results obtained if they are decomposed into smaller, less complex ones with clear instructions. 

\begin{mdframed}
    \textbf{Answer to RQ2:}
    A crowd of lay workers is capable of annotating online user feedback by requirements-relevance and even functional and quality aspects. Clear instructions help to convey the required expert knowledge. The best performance we attained was through a series of micro-tasks, each focusing on only a few classes.
\end{mdframed}

\vspace{1em}
    \begin{enumerate}[\bfseries RQ3.]
        \item \textbf{How to design an LLM pipeline so that it can achieve good results in classifying quality aspects in online user feedback with little training data?}
    \end{enumerate}
\noindent For RQ3, we performed a $2\times2\times2$ between-subjects experiment, with a pipeline that prompted two LLMs from the ChatGPT family, as well as the factors prompt type and learning strategy, resulting in eight conditions. Investigating how increasing task complexity affects the performance of LLMs (\textbf{RQ3a}), we found that ChatGPT achieved high precision but mostly limited recall for the binary classification task over online user reviews (Phase P1). On the somewhat more difficult binary classification task over individual sentences (P2), precision was still high, but recall worsened. The most difficult multiclass classification task over individual sentences (P3${^\prime}$) resulted in mediocre precision and recall. The performance of the LLMs worsened as task complexity increased, resulting in a decrease in the average accuracy scores of the conditions across the three phases of our experiment. The LLM used had the greatest impact on performance, as ChatGPT 4o made significantly better predictions on simpler tasks than ChatGPT 4 Legacy. However, ChatGPT 4o's upper hand diminished across the three phases, while the performance of ChatGPT 4 Legacy only gradually worsened. These findings suggest that LLMs are suitable but not perfect for this task.

Because some literature suggested that LLMs are likely to perform better with a specifically engineered prompt than with instructions not specifically engineered for LLMs~\citep[e.g.,][]{Brand23, ElHajjami24}, we studied how the use of engineered prompts and examples affects the performance of LLMs (\textbf{RQ3b}). In Phases P1 and P2, the prompt type did seem to have an influence on performance, but surprisingly it was a reverse effect. The Ky\={o}ryoku prompt not specifically engineered with and for ChatGPT tended to lead to better results than the specifically engineered one. This effect was significant for P1, and in P2 had a trend toward significance. In Phase P3${^\prime}$, the prompt type neither positively nor negatively affected performance. Possible explanations for this finding include our limited experience with prompt engineering despite the iterative approach taken to carefully align with ChatGPT, the shorter length of the Ky\={o}ryoku prompt allowing ChatGPT to find important guiding information more easily, or the notion that ChatGPT can be considered a layperson similar to a crowd worker, which was the intended target audience of the Ky\={o}ryoku instructions.

Regarding the role of learning strategy, the literature was inconclusive. In P1, there was an interaction effect for prompt type $\times$ learning strategy, suggesting that the examples we provided mostly had no effect on performance, except helping ChatGPT 4 Legacy to perform slightly better in P1.~\citet{Reynolds21} provide a good explanation on how to understand the role of few-shot examples. They argue that LLMs do not use prompts for learning, but instead to assign the appropriate task location in the LLM. Consequently, the few-shot examples merely steer the retrieval closer to or further away from the correct knowledge space. In our research, the definitions for each class might have been sufficiently self-explanatory for the examples to affect task location.
 
Regarding performance by class, we observed a class imbalance for ChatGPT 4 Legacy on P1, and for both LLMs on P2, which mostly overclassified items as \textit{helpful}. This pattern seems to resemble the phenomenon that ML-based classifiers applied to unseen data default to the positive answer when in doubt, prioritizing recall over precision. We are unsure why this happens when the mechanics are different; ML draws inferences from the data on which it was trained, while the decoder architecture of LLMs predicts the most likely response based on its model~\citep{Raffel20}. In our experiment, this entailed predicting which of the predefined class names should be returned. That ChatGPT 4o behaved differently between P1 and P2 might be due to it erring on the side of caution, for example, in P2 after sentence splitting due to the lack of context from a full review. For the multiclass classification in P3${^\prime}$, we found that both LLMs tended to overclassify \textit{feature request} and \textit{stability} and underclassify \textit{user-friendliness} and \textit{security}, resulting in low recall. The LLM conditions rarely achieved a balance between precision and recall for a single class. This suggests that on this complex task, both LLMs could not always correctly discern between the classes presented to them.

\begin{mdframed}
    \textbf{Answer to RQ3:}
    LLMs are capable of performing classification tasks by requirements-relevance and into functional and quality aspects. In our study, ChatGPT 4o mostly outperformed ChatGPT 4 Legacy, the choice of prompts sometimes improved performance, but providing examples (i.e., few-shot learning) had no effect.
\end{mdframed}

\vspace{1em}
    \begin{enumerate}[\bfseries RQ4.]
        \item \textbf{Which low-data approach is most suited for classifying quality aspects in online user feedback, considering their effectiveness and trade-offs?}
    \end{enumerate}
\noindent We both quantitatively and qualitatively compared the three approaches to identify the most suitable one for identifying quality aspects in online user feedback. The crowdsourced Ky\={o}ryoku method and the LLM pipeline achieved comparable results on the classification problem of annotating quality aspects in online user feedback, and were much more accurate than the LP-based approach. This finding highlights that when appropriate measures are taken, this complex classification task can be performed by a hybrid system (crowdsourcing) or an expert system (LLMs) instead of being performed or supervised by domain experts. 

Regarding the suitability for determining requirements-relevance (\textbf{RQ4a}), the crowd and the best-performing ChatGPT condition performed equally well on P1, with the crowd achieving better recall and ChatGPT achieving better precision on \textit{useless}. The crowd performed about equally well in Phase P2, but the best-performing ChatGPT condition performed worse, and ChatGPT's majority vote prediction was even less accurate, especially in terms of recall. This might suggest that human workers are better at considering sentences in isolation in P2, and that ChatGPT in part relied on the context provided in the complete online user reviews of P1, making it more difficult for ChatGPT to draw correct inferences if this context was missing. Interestingly, the best ChatGPT condition in P1 and P2 is $\lcurvyangle$Ky\={o},\,Few,\,4o$\rcurvyangle$, which happens to be the one prompted with the exact same instructions as the crowd workers received. Thus, ChatGPT 4o is capable of equating human workers on a classification task when given the same instructions, but the crowd was generally better at predicting whether online user feedback is requirements-relevant. 

Our analysis of how suitable each approach is to distinguish between the quality aspects (\textbf{RQ4b}) showed that classifying quality aspects in online user feedback remains a challenging task, given that it is a multiclass classification task based on an inherently arbitrary taxonomy of software qualities over intrinsically ambiguous online user feedback and performed without training data. The LP-based approach faces challenges due to its rigidity, but the approaches using crowdsourcing and LLMs also performed significantly worse on the classification task(s) in P3 \& P4, although they usually can correctly predict the quality characteristics with at least good precision or recall. Three conditions were equally accurate: the crowd in the P3$\rightarrow$P4 sequence, the best ChatGPT condition, and ChatGPT's majority vote prediction. The other two conditions in our comparison\textemdash the LP-based approach and the crowd in Phase P3${^\prime}$\textemdash did not perform well. The equal performance between ChatGPT and a crowd of human raters on this multiclass classification task suggests that the advantage of the knowledge processing capacity and training of an LLM does not necessarily make it superior to a crowd of lay workers who have typical human cognitive limitations when predicting quality aspects in online user feedback.

Due to the different patterns we observed, we found no conclusive evidence that a specific classifier is the best on a per-class level. Our classifier conditions showed patterns that resembled those found with traditional ML classifiers. For example, as in our study,~\citet{Kurtanovic17} obtained the highest precision and recall values for \textit{performance}, and the lowest for \textit{operability} and \textit{usability}, which in this study both belong to \textit{user-friendliness}. Although the best-performing configurations for crowd workers and LLMs performed equally well, we did determine that each approach tends to behave differently. For example, Table~\ref{tab:synth_confusion_p3} illustrates how the precision and recall by Crowd P3$\rightarrow$P4 was the inverse of ChatGPT Majority Vote (and, to a lesser extent, ChatGPT Best); while the former mostly excelled on recall and the latter mostly on precision, it was also the opposite for some other classes. 

We see further improvement potential for all approaches. The crowd will likely perform better if more judgments are collected, because a greater agreement between crowd workers has a higher accuracy than the agreement between LLM conditions. For the approach using LLM, a pipeline specifically designed for results based on a majority vote and the use of variations in prompts could further boost its accuracy. Even though the LP-based approach did not compare well to our other approaches, its ability to predict \textit{stability} well, and its potential for explainability and reusability might still be of use for settings that prioritize predictable outcomes and transparent processes, which the other approaches cannot provide due to the personal heuristics employed by the crowd workers and the non-deterministic nature of LLMs. The LP-based approach could therefore benefit from automating the process of curating the lexicon of keywords and vetting the LPs using ML approaches or LLMs. 

\begin{mdframed}
    \textbf{Answer to RQ4:}
    Crowdsourcing was overall better in determining requirements-relevance. Crowdsourcing, the best-performing ChatGPT condition and the majority vote prediction of ChatGPT classified quality aspects in online user feedback equally well, with no clear classifier being the best on individual classes.
\end{mdframed}

\section{Threats to Validity} \label{sec:ttv}
Several threats to the validity of our studies' results should be considered. We discuss the main validity threats according to the dimensions of construct, internal, external, and conclusion validity~\citep[cf.][]{Wohlin00}. 

\textit{Construct validity} is concerned with the appropriateness of the research approach and the instrumentation to answer the research questions. For RQ1, we used an empirically validated method for eliciting quality requirements. The use of other linguistic approaches might have led to different outcomes. The evolution of the elicitation workshops for RQ1 due to the adjustments we made to improve the effectiveness of the instructions and templates could have changed the group dynamics, but these changes were minor. Although we had to switch from co-present to remote workshops due to the COVID-19 pandemic, we have not observed an impact on the workshops' dynamics or outcomes. We made the best possible effort to reduce bias and misinterpretation regarding quality characteristics, especially by dedicating time at the beginning of the workshop to making the participants' preexisting notions explicit, and to compare their understanding with each other and with the ISO 25010 definition. 

For RQ2,~\citet{Vliet20} discusses the threats to validity in the design of Ky\={o}ryoku, which particularly focused on the lack of reference material and practical experience with crowdsourcing tasks. The finding that a sequence of micro-tasks leads to better results is based only on the P3$\rightarrow$P4 sequence and would have been stronger if we had considered multiple sequences in which we varied the composition of classes. 

The main discussion of construct validity concerns RQ3, because from an empirical point of view, our understanding of LLMs is still limited and they produce output that is non-deterministic. Several threats pertaining to study participants apply to measuring the behavior of LLMs as a generative entity, which in this work may include history, maturation, and testing effects, but in particular concerns the \textit{multiple-treatment interference}, which~\citet{Campbell63} describe as a threat through which generalizability is reduced when applying multiple treatments to the same respondents whenever the effects of prior treatments are not erasable. We actively mitigated the dependency of messages within a single ChatGPT conversation by performing all the prompt sessions separately, with ChatGPT's memory function disabled. 

The choice to use ChatGPT in research is not undisputed, as it has both clear advantages and marked disadvantages. The purpose of this research was to investigate the best result that LLMs could potentially achieve at the time of the experiment; not an exhaustive comparative study between LLMs. Based on an extensive pretest~\citep[see the online appendix;][]{appendix}, we determined that ChatGPT was the most-researched and purportedly the most robust LLM for achieving that goal. From among the alternatives, it was the most capable in handling our prompts and producing meaningful results. We encourage the evaluation of other LLMs through reproductions of this experiment. However, it is appropriate to elaborate on its disadvantages. We chose ChatGPT over prompting GPT through an API because we had established that it had just received the feature of handling Excel files and that it was capable of processing larger amounts of text. In retrospect, using GPT might have been less time consuming due to the amount of troubleshooting necessary, and it could also have allowed certain hyper-parameter settings such as temperature or top-$k$ to be altered, potentially affecting classification accuracy. We do not expect that the use of the Web interface affected our ability to program the experiment or to avoid errors, but we did encounter response limits followed by cooldown periods, and we could not control the number of tokens, which is why we reported on the number of input characters instead. The use of ChatGPT is dependent on design decisions taken by its developer OpenAI; its performance can be unpredictable due to its opaque evolution that in turn also impairs its replication potential, but we also observed improvements being made to its processing speed. We estimated this limitation to be acceptable considering the response quality we established for the two state-of-the-art LLMs we compared. We consider that the two additional experimental factors of prompt type and learning strategy that we chose based on our pretest were appropriate for our comparison purposes. 

Finally, LLMs are under scrutiny from an ethical stance, which includes concerns about regulations and data privacy, transparency, hallucinations, and harmfulness~\citep{Vogelsang24}. These should be considered in the context of the other approaches considered. Although our dataset was curated from publicly available sources, we ensured that our data was anonymized~\citep{Groen19,Vliet20}, and fixed response options limited the potential for hallucination. The harmfulness particularly pertains to sustainability concerns about its energy consumption. Here, we note that LLMs require a surge of energy, while our other approaches rely on power consumption over a prolonged period of time for vetting and deploying LPs or for performing the decentralized crowd work on hundreds of devices. Similarly, one should consider the trade-off between whether it is desirable to strain an expert with a cognitively demanding task, whether voluntary crowd workers from impoverished countries are exploited or if it allows them to earn additional income~\citep[cf.][]{Sachs05}, and whether or not this outweighs sponsoring the shareholders of the big tech companies operating LLMs.

\textit{Internal validity} addresses the probability that the same results will be obtained if a study is to be repeated. For RQ1, the variation between experts could cause workshops with different expert participants to produce different results. Nevertheless, this likely has only a limited effect on our interpretation of how well LPs can predict quality characteristics in online user feedback. Here, the primary weakness of using quality-related (combinations of) keywords recoded into LPs is the low recall that we were able to achieve. Because only predefined words were found, the LPs were inherently prone to missing relevant keywords that were not elicited. We previously established that an approach in which keywords are extracted from an existing dataset can only partially mitigate this~\citep[see Study II in][]{Groen17users}, which for this work would have been a too large deviation from the NFR Method, and it would have interfered with our ability to draw inferences about expert contributions. In RQ2, we particularly saw that the crowd we had attracted for the tasks of P4 and P3${^\prime}$ was different compared to P1, P2, and P3. Especially the fairness of remuneration was rated lower, even though we paid more. The crowd workers also found the tasks to be more difficult, even though P4 should have been easier than P3. This might suggest that crowd workers attracted at different points in time might have different results. The design of the approach in which each phase utilized input from the preceding phases is likely to perpetuated errors further; especially items not presented in P4 that the crowd workers in P3 failed to classify as \textit{quality}. Despite this, the two consecutive Phases P3$\rightarrow$P4 did still achieve better results than Phase P3${^\prime}$. The quality of the results could possibly be improved by involving more crowd workers, although this also increases costs. The main threat to the internal validity of RQ3 is the non-deterministic nature of LLMs, the impact of which we discussed as part of the construct validity. Although within our study, we found that performance was consistent across conditions\textemdash despite different experimental factors\textemdash their behavior might vary at different points in time.

An important determinant for the internal validity of all three studies is our dataset. Due to the rapid evolution of the app landscape, the results may differ if more recent online user feedback had been used. However, we cannot gauge the impact of the age of the dataset because to our knowledge, no works have replicated or updated the work by~\citet{Pagano13} that could give an indication of whether any characteristics of the content of online user feedback have shifted over time. Based on the analysis of the confusion matrices, we have learned that more time needs to be spent on discussing the gold standards, both by reconciling differences in tagging and also in verifying the consistency. In~\citet{Vliet20}, this was partially mitigated by manually inspecting the differences and adding a lenient judgment. To avoid this unnecessary step, for this paper we refined the gold standards through iterative sessions between three authors for a more reliable comparison. Due to the choice of a bootstrapped random sample from a larger dataset, the different classes had varying numbers of sentences associated with them, resulting in the strength of the performance calculations to vary. This particularly affected \textit{security}, which had the lowest number of sentences. We also performed extensive manual sanitation of the data upon finding contamination issues, in order to achieve a fair and reliable analysis. This resulted in the omission of approximately 15\% of the data for RQ2, especially in P2, which could have negatively affected the performance values of the crowd. 

Regarding RQ4, to ensure comparability between approaches, we had to make compromises, particularly in terms of their classification. The LP-based approach was limited to only one label, and to keep the micro-tasks in Ky\={o}ryoku simple, each crowd member could provide just one label, from which a multi-label prediction arose in case of a tie in the majority vote. We instructed ChatGPT to assign a multi-label response only in case of a tie between possible labels, and ties in the majority vote prediction also resulted in multi-label predictions. We considered the multi-label items in the gold standard for P3 \& P4 as a range of correct options, only one of which needed to be matched for a correct response. We leave explorations of multi-label classifications and attributing classes to smaller portions of the input text to further research. In addition, each approach had unique advantages and disadvantages that could affect performance, complicating the objective comparison of their strengths and challenges. The LP-based approach benefited from iterative refinements made to the LPs, of which the high precision could not be offset at that stage against objective recall scores; in the Ky\={o}ryoku method, the crowd achieved good scores in Phase P4 due to the smaller number of classes, and ChatGPT benefited from having eight experimental conditions, as opposed to one for the LP-based approach, and two in Ky\={o}ryoku (P3$\rightarrow$P4 and P3${^\prime}$). To compensate for the weaknesses of each approach, in the Ky\={o}ryoku method, we compensated for the mental limitations of laypeople by breaking down the task into multiple classification stages; we provided the LLMs with more context and engineered prompts to prevent potential misinterpretations, and we iteratively refined and optimized the LPs. 

\textit{External validity} concerns the generalizability of the results. In RQ1, the workshops were conducted only once for each quality characteristic, but we argue that the discussion of the experts participating in the workshops sufficed to produce a well-composed keyword meta-model. The workshop participants were not frequent authors of online user reviews themselves, but a pretest had shown that computer laypeople provided fewer relevant keywords \& phrases, and~\citet{Tizard20} found that typical authors of user reviews have greater computer affinity, which provides some evidence that our participants can be considered sufficiently representative. In addition, or work shows similarities with other research in this direction, such as~\citet{ClelandHuang07}, who found that for several indicator words, it is the context that prescribes the subcharacteristic to which it should be attributed, and that the challenge accordingly lies in the boundaries between the quality characteristics. For example, do user reports about being able to do something with the app really ``fast'' relate to the user's perception of the app's \textit{time behavior} or its \textit{operability} in terms of the workflow? 

In RQ2 and RQ3, the main threats to the external validity are a logical consequence to the already described changes in the composition of the attracted crowd or of the LLM used at a given point in time. The majority vote prediction might be more robust against changes in the performance of individual crowd workers or ChatGPT conditions. Based on the results of the pretest we performed for RQ3, we consider it unlikely that the pipeline we created will lead to the same results when prompting other LLMs than those tested.

\textit{Conclusion validity} pertains to the credibility of the results obtained. We have done our best to ensure the quality of our data and interpret it using tabulations, visualizations, and appropriate statistical measures. Due to the newness of LLM technology, we have attempted to base our decisions on how to analyze it based on the available literature on evaluating classifiers, as part of which~\citet{DellAnna23} is recognized as authoritative work~\citep[cf.][]{Vogelsang24}. However, during this process, we faced important fundamental questions, such as whether the same LLM in a different condition should be analyzed using a repeated-measures test, or if its non-deterministic nature is a reason not to do so. Due to the novelty of this domain, best practices for evaluating LLMs have not yet been developed, and these might favor other metrics and tests than the ones we chose. We share our raw data in the online appendix~\citep{appendix}, over which alternative statistical tests could be performed.

\section{Related Work} \label{sec:rw}
We discuss the main related work by addressing crowdsourcing in RE (Section~\ref{sec:rw-cs}), the automated analysis of online user feedback (Section~\ref{sec:rw-nlp}), and LLMs in RE (Section~\ref{sec:rw-llms}). 

\newpage

\subsection{Crowdsourcing in RE} \label{sec:rw-cs}
The term \textit{crowdsourcing} originated from the business sector, when~\citet{Howe06} coined it as ``the act of taking a job traditionally performed by a designated agent (usually an employee) and outsourcing it to an undefined, generally large group of people in the form of an open call''. In his comprehensive overview,~\citet{Howe10} refers to this definition as ``the White Paper Version''. In this traditional perception of crowdsourcing, a crowdsourcing platform offers a crowdsourcing task of a crowdsourcer to a crowd of workers~\citep{Hosseini14pillars,Hosseini15taxonomy}. Our Ky\={o}ryoku method used for RQ2 belongs to the body of work that approaches crowdsourcing in this way. 
Expanding on his own definition,~\citet{Howe10} provided an additional perception of crowdsourcing, where it is seen as ``the application of Open Source principles to fields outside of software''. Howe refers to this pragmatic perception of crowdsourcing as ``the Soundbyte Version''. No known work in RE specifically addresses the pragmatic perception, but it illustrates that there can be more than one notion of crowdsourcing. 
The majority of works in RE in fact assume what could be described as a modernist perception of crowdsourcing, which considers crowdsourcing to involve any approach where information is obtained (i.e., ``sourced'') from a crowd, typically not bound by Open Source principles and usually for profit. Neither Howe nor the literature provide an official definition for this view.~\citet{Wang19} provide an overview of approaches in RE that make use of ``crowdsourced user feedback'', where crowdsourcing encompasses obtaining (i.e., ``sourcing'') requirements from the crowd’s online user feedback. In RE, this is typically referred to using the umbrella term \textit{Crowd-based RE}, which encompasses all approaches to (semi-)automatically analyze and classify online user feedback for the specification and evolution of software products in RE~\citep[CrowdRE;][]{Groen15refsq,Groen17ieee,Glinz19}. Many of the approaches suggested in CrowdRE focus on the analysis of natural-language texts. Engaging the crowd mainly relies on a form of self-regulated social participation~\citep{Ali12}. We consider all three approaches proposed in this work to fall under the CrowdRE umbrella.

The early 2010s saw a drastic rise in the interest to adopt traditional crowdsourcing in many domains including SE~\citep[see][for reviews]{Mao17,Ghezzi18}. However, reports of poor results from crowdsourced activities might have caused the upswing to lose traction. Notably,~\citet{Stol14} presented a case study in which software programming work was crowdsourced, where none of the promises of crowdsourcing\textemdash cost reduction, faster time-to-market, high quality, and creativity \& open innovation\textemdash were attained~\citep{Fitzgerald18}. Their work shows that crowdsourcing might not be suited for tasks that are more complex or require coordination between developers.

Regarding crowdsourcing for RE, in their seminal work on this topic,~\citet{Hosseini14towards} discuss five crowdsourcing approaches in RE, for which they suggested appropriate crowdsourcing platforms to be developed: requirements-driven social adaptation, feedback-based RE (i.e., CrowdRE), stakeholders' discovery, requirements identification, and empirical validation. They moreover discuss ten crowdsourcing features and their associated quality attributes, which were expanded upon in~\citet{Hosseini15configuring} with a review of the challenges faced for each of these features.~\citet{Khan21} performed an interview study with 50 practitioners to elicit challenges when using crowdsourcing platforms in RE, finding challenges related to time, quality of work, culture, number of experts, communication and confidentiality, conformity between feedback and requirements, and degree of knowledge on crowdsourcing. 

Among the tools proposed to use crowdsourcing in RE is \textit{CrowdREquire}~\citep{Adepetu12}, an envisioned online platform to crowdsource requirements specification in the form of a competition. Most other works that involve traditional crowdsourcing in RE tasks have been found to particularly focus on mechanisms involving non-RE actors in requirements elicitation and validation~\citep{Groen15crowdout,Hosseini14pillars,Hosseini15configuring}. For example, \textit{StakeSource}~\citep{Lim10} is a research-based commercial crowdsourcing platform that supports stakeholder selection through peer recommendations and involves them in requirements elicitation and prioritization. It in part draws from the \textit{Organizer \& Promoter of Collaborative Ideas}~\citep[OPCI;][]{Castro09}, where stakeholders could co-create requirements in a forum that automatically clustered related ideas and used a recommender system to link stakeholders to potentially relevant threads. The \textit{Crowd Requirement Rating Technique}~\citep[CrowdReRaT;][]{Oluwatofunmi20} was designed to crowdsource requirement ideation and rating, and a tool developed by~\citet{Sari21} proposed the use of crowd discussions in the elicitation process. More recent efforts focus on identifying and selecting suitable stakeholders through crowdsourcing~\citep{Alamer23,Delima23}.

Although the traditionalist perception of crowdsourcing does not include indirectly obtaining online user feedback, several approaches in this line of research suggested ways to actively involve crowd workers in a crowdsourced process to solicit online user feedback, perform feedback analysis, or validate the results~\citep{Groen15crowdout}.~\citet{Snijders15} proposed \textit{REfine}, a gamified CrowdRE approach that involves crowd members in gathering needs and in the subsequent requirements analysis and prioritization. This work was extended in the \textit{Crowd-based Requirements Elicitation with User Stories} method~\citep[CREUS;][]{Wouters22}, which gathers ideas from the crowd using the user story notation through four phases: preparation, idea generation, refinement, and response and execution. The initial ideas presented for \textit{Requirements Bazaar} envisioned a similar solution for eliciting requirements through a platform, but additionally included considerations for requirements negotiation~\citep{Renzel13}. The \textit{Crowd-Annotated Feedback Technique}~\citep[CRAFT;][]{Hosseini17craft} crowdsources the classification of app store feedback according to RE-related dimensions. Crowd members act as taggers, who select a particular fragment of online user feedback, pick a category and a (sub-)classification, and provide an assessment of the feedback quality and one's own confidence level. Although CRAFT was designed as an in-situ mechanism and did not offer remuneration as an incentive, its validation demonstrated the possibility of outsourcing online user feedback annotation to a crowd, which we later built on when designing \textit{Ky\={o}ryoku} to crowdsource online user feedback classification through micro-tasks~\citep[][see RQ2]{Vliet20}, which is the basis for RQ2. Aside from our work, the only other example of research using an RE-related crowdsourced annotation task is~\citet{Stanik19a}, who based part of their ML and DL training set on 10,000 English and 15,000 Italian tweets annotated by the crowd. Our work in RQ2 focuses on how this annotation task can be optimally configured.

\subsection{Online User Feedback Analysis in RE} \label{sec:rw-nlp}
In RE, natural language processing (NLP) has mainly been concerned with identifying content in documents for RE tasks~\citep{Ferrari25}. In our work, we have applied a linguistic approach for RQ1, and we performed requirements classification with all three approaches considered.

\textit{Linguistic approaches.} 
Performing text mining using a linguistic approach involves identifying patterns in utterances that point to a possible use or intention~\citep{Jurafsky09}. Although most of the recent literature has favored the use of traditional ML~\citep[see][for a review]{Santos19approaches}, linguistic approaches such as keyword tracing have successfully identified requirements in natural language texts. These are rooted in strategies that long preceded the evolution of online user feedback analysis.~\citet{Cysneiros01b} proposed a strategy to integrate NFRs into conceptual models by constructing a Language Extended Lexicon consisting of a shared application vocabulary that links representations like requirements and models together. This technique could be used to elicit, analyze, and trace NFRs. This makes domain ontologies suitable for requirements management activities such as glossary development or trace link vetting~\citep{Dermeval16}. For example, the \textit{OpenReq Dependency Detection} tool~\citep[OpenReq-DD;][]{Motger19} automatically generates ontologies and clusters keywords to detect requirements dependencies~\citep{Deshpande20}. For RQ1, we developed a keyword meta-model for software product quality characteristics in the form of a lexicon.

The first application of linguistic techniques to online user feedback analysis was the \textit{Mobile App Repository Analyzer}~\citep[MARA;][]{Iacob13}, which identifies feature requests in text fragments of user reviews using language patterns and aggregates results. Regular expressions have proven useful for extracting positive feedback, negative feedback, and feature requests from user reviews~\citep{Fu13,Iacob13}.~\citet{Panichella15} derived 246 \textit{recurring linguistic patterns} through a manual inspection of 500 user reviews. These patterns describe natural language heuristics that indicate syntax structures that suggest the presence of feature requests, for example: \code{``[something] needs to be [verb]''}. In turn, they used a parser that labels grammatical relationships between words within a sentence. This approach achieved fair precision and recall. The research group continued to develop the \textit{User Request Referencer}~\citep[URR;][]{Ciurumelea17}, which classifies reviews according to a custom taxonomy, and then performs pattern matching using Information Retrieval techniques to identify the affected source code artifacts. In~\citet{Groen17users}, we theorized that online user feedback will often contain redundancies. We obtained words specifically pertaining to \textit{usability} from 360 user reviews and translated these into 16 linguistic patterns, achieving good precision\textemdash 0.92 on average\textemdash but presumably limited recall.~\citet{Paech20} and~\citet{Schrieber21} present research on the \textit{User View Language}, an intermediate language with textual and visual notation of a software's outside view that aims to improve user--developer communication. Their work asserts that users have a different mental model than a system's developers, which will cause descriptions of similar context to be described differently in utterances such as online user feedback. Their research suggests that the user view is quite consistent, regardless of technical background~\citep{Anders22} or the frequency of using an app~\citep{Anders23}. In spite of these works, we found that a comprehensive investigation of how suitable linguistic patterns are for user feedback analysis was missing. In RQ1, we used regular expressions to identify the end-user view on software quality.

\textit{Requirements classification.} 
The proposed solutions for classification problems into RE dimensions for structured or unstructured documents have most often employed NLP techniques backed by traditional ML models~\citep[see, e.g.,][for reviews]{Santos19approaches,Santos19taxonomy,Wang19,Khan19}. This section focuses on research efforts of classifying dimensions of non-functional requirements or quality requirements, because this is the main focus of our work. 

Before online user feedback was considered a viable source of information for RE,~\citet{ClelandHuang06} used an Information Retrieval approach with a predefined fixed set of keywords extracted from standardized catalogs to detect and automatically classify NFRs in requirements specifications. This was extended with early demonstrations of using speech recognition to classify NFRs in meeting and interview recordings~\citep{Steele06}, and to analyze stakeholders' quality concerns from unstructured documents such as meeting minutes, interview notes, and memos~\citep{ClelandHuang07}, with the goal of reducing the number of overlooked NFRs through manual discovery. The work of~\citet{ClelandHuang07} bears the greatest resemblance to our work on RQ1. They created keyword taxonomies for \textit{security} and \textit{performance}, which they then used to iteratively retrain a classification algorithm to detect and classify (i.e., predict) NFRs through a series of experiments performed over 30 documents, increasing its recall from 62.6\% to 79.9\%, and its precision from 14.7\% to 20.7\%. In RQ1, we detected and classified the quality concerns that crowd members express in online user feedback in a similar way. 

NFRs are usually found in technical documents such as requirements specifications or SE contracts, which are typically well-curated and validated product requirement datasets. The analysis of such documents has become a popular field of study, and in recent years, techniques based predominantly on DL have been proposed to classify NFRs and other requirements~\citep[e.g.,][]{Navarro17,Tamai18,Hey20,Sainani20,Ajagbe22}. The performance of these classifiers is most often validated on the \textit{PROMISE} dataset~\citep{Sayyad05,ClelandHuang07data} of user requirements. However, they face limitations when they are applied to online user feedback~\citep{Reddy21, Dabrowski22}. CrowdRE research has demonstrated the suitability of online user feedback to obtain quality requirements from~\citep[e.g.,][]{Groen17users,Jha17,Lu17}, but unlike structured technical documents, online user feedback consists of unstructured, noisy, and inherently ambiguous texts not specifically intended for RE. It does not readily contain existing NFRs, which makes it more difficult to analyze. In their review,~\citet{Dabrowski22} found that the analysis of NFRs comprises one of the three central use cases in online user feedback analysis. Through automated identification, classification, summarization, and sentiment analysis, NFRs can be elicited, and problems can be identified, categorized into quality attributes, specified ad hoc with a documented rationale, and prioritized. Research works addressing the problem of (systematically) discovering quality requirements with classification methods have typically proposed methods employing traditional ML or DL~\citep{Zhou14,Yang15,Groen17users,Lu17,Jha17,Kurtanovic17,Williams17,Stanik19a,Reddy21}. Although ML-based text mining tools in RE can achieve high recall values, their main weakness is usually that their precision is often lacking~\citep{Jha17}. This is why we investigate the suitability of other approaches for this classification problem.

\subsection{LLMs in RE}  \label{sec:rw-llms}
In RQ3, we consider the suitability of LLMs by example of ChatGPT to classify online user feedback according to RE-relevant dimensions. Due to the novelty of LLMs, little research on their application in RE had been published when we performed our investigation. A growing body of work in RE is considering the role of LLMs in education~\citep{Abdelfattah23, Carvallo23}, and explored the generation of requirements~\citep{Brand23, Bencheikh23}, models~\citep{Ruan23, Arulmohan23,Bragilovski24} or artifacts such as personas or interview scripts~\citep{Gorer23, Brand23}.

Regarding the classification using LLMs, like in RQ3, in the seminal work of~\citet{Gilardi23}, ChatGPT using a zero-shot prompting strategy was found to have outperformed human raters on classifying tweets and news articles in a political sciences study. These were classified, among other things, by relevance, topic, sentiment, and stance. The only negative effect of the zero-shot prompting strategy might have been its diminished ability to determine the relevance of some texts due to a lack of examples, and~\citet{Gilardi23} suggest that future work should consider few-shot prompting strategies, citing evidence that this strategy is best suited for LLMs~\citep{Brown20}. Although the domain is different, their work shows some similarities with our crowdsourced task in RQ2 and the use of this data in a study with LLMs in RQ3, which also classified the data according to relevance (Phases P1 \& P2) and aspects that are similar to topics (Phases P3 \& P4). Particularly their topic annotation task resembles our Phase 3${^\prime}$ by complexity.

So far, only few known works have investigated automated classification according to requirements-relevant content using LLMs as in our RQ3.~\citet{ElHajjami24} found that their ChatGPT models (3.5 Legacy [DaVinci], 3.5 Latest [Turbo], and 4) almost always outperformed LSTM- and SVM-based classifiers in binary classifications replicating those of~\citet{Dalpiaz19b}. Their conflicting results on the use of examples might suggest that these worsened the task location of the LLM~\cite[cf.][]{Reynolds21} or introduced bias. Recent works have also investigated the classification of requirements with learning.~\citet{Wang24} compared the ability of various language model types to predict requirements in two preexisting annotated datasets, and found DeBERTa to outperform Llama2 and RoBERTa, and that ensembles can achieve better results.~\citet{Can25} employed LLMs to identify and then classify requirements in backlog items in issue tracking systems. They found that the encoder-only models they used, BERT and RoBERTa, outperformed the decoder-only models, ChatGPT 4 with and without in-context learning, Mistral 7B, and Llama 3.

At the time we started our research, little work specifically on prompt engineering existed or was known to us. We found helpful guidelines in~\citet{Pattyn23} and~\citet{ElHajjami24}, who suggested that the input data, context, procedural guidance classification schema, and expected structure of the output must be clearly described in the prompt syntax. In her practice report,~\citet{Brand23} furthermore determined that instructions should be clear, specific, and written in a way that triggers the model to provide versatile responses. More useful resources are now available, including a catalog of prompt patterns~\citep{White23}, the \textit{OpenPrompt} framework~\citep{Ding22} for prompt learning and prompt management for research, the Automatic Prompt Engineer method~\citep[APE;][]{Zhou23} for automatically composing and selecting instructions, and datasets provided by~\citet{Das23} including one to evaluate zero-shot models through Natural Language Inference.

\section{Conclusion \& Future Work} \label{sec:conclusion}
Identifying software product quality characteristics in inherently ambiguous online user feedback is a complex classification problem, consisting of a multiclass and possibly multi-label classification according to some taxonomy such as ISO 25010~\citep{ISO11}. The automation potential for this classification problem is inhibited due to the lack of large training corpora, resulting in a low-data environment, which limits the potential for automated classifiers such as traditional ML or DL algorithms to be trained and perform this task with high accuracy. This is why in this research we investigated and compared three low-data approaches, in order to answer our main research question (\textbf{MRQ}): ``Which low-data approach is most accurate in identifying quality aspects in online user feedback?''

We decomposed our MRQ into four research questions (RQs) to analyze and compare these approaches in their analysis of a sample of 1,000 app store reviews. For \textbf{RQ1}, we analyzed traditional linguistic approaches through an adaptation of the NFR Method, through which we obtained a set of language patterns (LPs). Although it successfully led to a lexicon of keywords and the creation of a set of LPs, the vetting process was intensive work and an experiment showed that the classification only predicted the quality characteristic of \textit{reliability} sufficiently well. For \textbf{RQ2}, we extended prior research on the Ky\={o}ryoku method~\citep{Vliet20} to crowdsource the classification of quality aspects in online user feedback. An experiment showed that a series of smaller classification tasks achieves better results than a single, larger classification task. It demonstrated that it is possible for online user feedback to be classified by a crowd of lay workers into all quality aspects. For \textbf{RQ3}, we modeled an LLM pipeline after the structure used in RQ2 to determine its ability to achieve good results on this classification task with little training data. An experiment showed that ChatGPT 4o outperformed ChatGPT 4 Legacy in classifying by requirements relevance, and that ChatGPT 4o and the instructions that we provided to the crowd in Ky\={o}ryoku rather than the specifically engineered prompt somewhat boosted performance. It suggests that LLMs are capable of being used for this classification problem. 

We performed a comparison of these three approaches in \textbf{RQ4} about which low-data approach is most suited for classifying quality aspects in online user feedback, considering their effectiveness and trade-offs. Our results showed that the use of crowdsourcing through the Ky\={o}ryoku method has a slight advantage over LLMs for classification by requirements relevance, but that the best configurations of both approaches proved to be equally suitable for classifying quality aspects, while the LP-based approach fell short. This suggests that both crowdsourcing and LLMs should be considered for \textit{complex tasks} in RE, especially for purposes for which there is not enough data available to train ML classifiers. Specifically, the Ky\={o}ryoku method\textemdash introduced in~\citet{Vliet20} and extended in this work\textemdash and our LLM pipeline with the associated artifacts can support the identification and classification of RE-relevant content in online user feedback. The implication of this finding is that the body of work on CrowdRE is extended with two suitable approaches for this purpose.

In this work, we have considered the three low-data approaches in isolation. A primary focus in future work should be to investigate whether a combination of approaches can help yield better results by compensating for each other's weaknesses and limitations. Ensemble methods for automatically classifying app reviews have previously been shown to perform better than individually applied techniques~\citep{Guzman15}. We see several possibilities for how these approaches can support one another. The effective use of LLMs depends on their collaboration with human operators, for example, supervision in learning. As a result, the best performance might be attained through a pipeline that, for example, uses automated analysis to verify the results from crowdsourcing and reattributes sentences for sequential phases. A keyword meta-model could help to expand the domain knowledge of the LLM. And the output of a crowd or an LLM could be grouped through linguistic analyses to find inconsistencies in the heuristics applied when tagging, with the LLM in turn suggesting potential to relabel the data accordingly. Moreover, given that the three low-data approaches do not provide in an algorithm that can be continuously trained to improve, we moreover see potential in the use of any of the approaches to construct a larger data corpus of online user feedback classified into quality aspects, which can in turn be used to train traditional ML and DL algorithms.

\section*{Acknowledgment}
The authors would like to thank all the participants of the keywords \& phrases elicitation workshops for RQ1, and Svenja Polst and Shivaji Yadav for their support in moderating the workshops. The authors thank Jonathan Ullrich for his insightful advice on RQ3. This research did not receive specific grants from funding agencies in the public, commercial, or not-for-profit sectors.
\printcredits

\bibliographystyle{apalike} 

\bibliography{main}

\end{document}